\documentclass[amsfonts, amssymb, amsmath, reprint, showkeys, twoside, prl, floatfix]{revtex4-2}
\usepackage[english]{babel}

\usepackage[utf8]{inputenc}
\usepackage[T1]{fontenc}

\usepackage{amsthm}
\usepackage{mathtools}
\usepackage{units}

\usepackage{xcolor}
\usepackage[pdftex]{graphicx}

\usepackage{hyperref}
\hypersetup{colorlinks=true, citecolor=blue, urlcolor=blue, linkcolor=blue}

\newcommand{\bra}[1]{\langle#1\rvert} % Bra
\newcommand{\ket}[1]{\lvert#1\rangle} % Ket
\newcommand{\expect}[1]{ \langle #1 \rangle} % Expectation value
\newcommand{\operator}[1]{\hat{#1}}

\newcommand{\mat}[1]{\mathsf{#1}}

\newcommand{\cone}{\mathrm{i}}

\newcommand{\edge}[1]{\ensuremath{\mathtt{#1}}}
\renewcommand{\vec}[1]{\boldsymbol{#1}}

\begin{document}
\title{Ultrafast Orbital Hall Effect in Metallic Nanoribbons}

\author{Oliver Busch}
\email[Correspondence email address: ]{oliver.busch@physik.uni-halle.de}
\author{Franziska Ziolkowski}
\author{B{\"o}rge G{\"o}bel}
\author{Ingrid Mertig}
\author{J{\"u}rgen Henk}
\affiliation{Institut f{\"u}r Physik, Martin Luther University Halle-Wittenberg, 06099 Halle, Germany}

\date{\today}

\begin{abstract}
The orbital Hall effect  can generate currents of angular momentum more efficiently than the spin Hall effect in most metals. However, so far, it has only been understood as a steady state phenomenon. In this theoretical study, the orbital Hall effect is extended into the time domain. We investigate the orbital angular momenta and their currents induced by a femtosecond laser pulse in a Cu nanoribbon. Our numerical simulations provide detailed insights into the laser-driven electron dynamics on ultrashort timescales with atomic resolution. The ultrafast orbital Hall effect described in this work is consistent with the familiar pictorial representation of the static orbital Hall effect, but we also find pronounced differences between physical quantities that carry orbital angular momentum and those that carry charge. For example, there are deviations in the time series of the respective currents. This study lays the foundations for investigating ultrafast Hall effects in confined metallic systems.
\end{abstract}


\maketitle

\paragraph{Introduction.} The concepts and ideas of spintronics have recently been expanded in two promising ways. First, in addition to exploiting the spin degree of freedom of electrons, their orbital degree of freedom is also used, leading to the field of orbitronics~\cite{cao2020, go2021, rappoport2023}. As a result, well-established effects such as the spin Hall~\cite{dyakonov1971, hirsch1999, kato2004, sinova2015} and the (spin) Edelstein effect~\cite{Aronov1989, edelstein1990, inoue2003, gambardella2011} are complemented by their orbital counterparts, namely the orbital Hall effect (OHE)~\cite{zhang2005, bernevig2005, kontani2008, tanaka2008, kontani2009} and the orbital Edelstein effect~\cite{zhong2016, yoda2018, salemi2019, johansson2021}.

Secondly, well-known steady-state phenomena are brought into the time domain by driving systems on the picosecond timescale (e.g., using terahertz radiation~\cite{hafez2016}) or on the femtosecond timescale (e.g., using ultrashort laser pulses~\cite{phillips2015}).

In this Paper, we aim to investigate theoretically the ultrafast orbital Hall effect (UOHE) in a metallic nanoribbon, thereby reuniting these two paths. As a result, we illuminate the interplay of laser-induced longitudinal and transversal charge currents, occupation dynamics, and accumulated orbital angular momentum (OAM), as well as OAM currents on the femtosecond timescale with atomic resolution.

\begin{figure}
    \centering
    \includegraphics[width = \columnwidth]{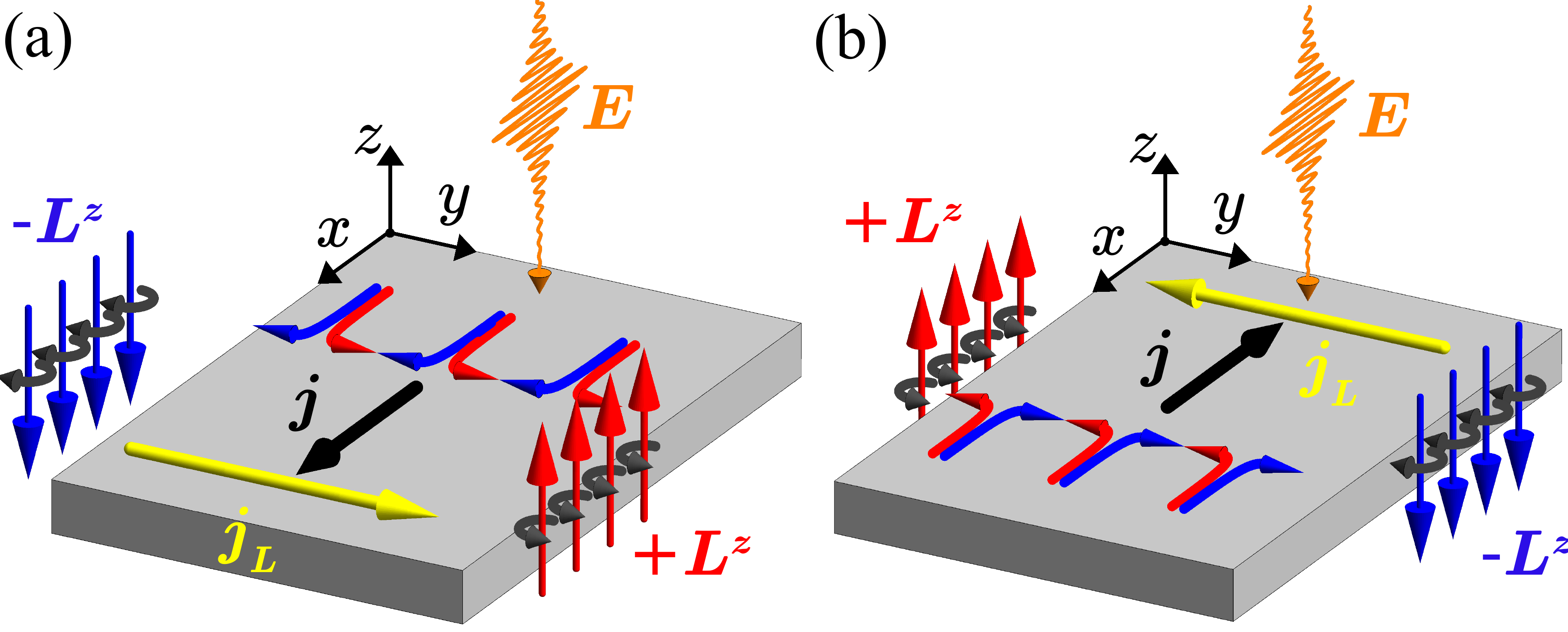}
    \caption{Snapshots of the ultrafast orbital Hall effect in a two-dimensional sample (gray rectangular solid). (a) A linearly polarized femtosecond laser pulse impinges perpendicular to the surface (along the $z$-axis) onto the sample. The laser's electric field $E$ (orange), oscillating along the nanoribbon ($\pm x$-direction), causes an oscillating longitudinal charge current $j$, which -- at the moment depicted here -- is oriented in $+x$-direction (black arrow) and is deflected toward the ribbon's edges; confer the three representative pairs of current filaments 
    (bent blue and red arrows). Hence, orbital angular momentum (OAM) $L^{z}$ is transported across the ribbon, giving rise to a transverse OAM current $j_{L}$ (yellow arrow along $+y$-direction). As a result, $L^{z}$ is accumulated with opposite orientation at the edges (upward red and downward blue arrows). (b) Half a laser's period later, the reversal of $E$ reverses the orientation of $j$, $j_{L}$, and $L^{z}$. The periodic field switching creates an ultrafast (on the femtosecond scale) orbital Hall effect.}
    \label{fig:sketch}
\end{figure}

The steady state is commonly addressed by a conductivity tensor, often calculated within linear response theory (e.g., Refs.~\onlinecite{seemann2015, nagaosa2010, sinova2015, go2018}). Such a tensor describes the relevant physics, but lacks details on the microscopic level; for example, it does not facilitate access to spatial resolution. Approaches relying on response tensors have been successfully brought into the time (frequency) domain. Examples include spin polarization and photocurrents calculated in dependence of frequency~\cite{dejuan2017, rostami2018, fregoso2019, sipe2000, xu2021, adamantopoulos2022}. However, the consideration of ultrafast transport phenomena in small samples requires a strategy that provides easy access to both spatial and temporal resolution.

In this Paper, we provide detailed insights into the electron dynamics that manifest themselves in the ultrafast orbital Hall effect (sketched in Fig.~\ref{fig:sketch}). We achieve disentanglement of the currents and the orbital angular momenta induced by a femtosecond laser pulse in a nanoribbon~\cite{mancini2015} with respect to space and time through numerical simulations performed within our theoretical framework \textsc{evolve}~\cite{Toepler2021, Ziolkowski23, busch2023a, busch2023b}. We show that the typical picture of the steady OHE~\cite{bernevig2005, kontani2008, tanaka2008, kontani2009, go2021} can be extended toward the femtosecond scale; it is thus appropriate for the UOHE in general. However, the rather strong perturbation of the system by the laser pulse attenuates the coherence among the involved physical quantities (depicted in Fig.~\ref{fig:sketch}) right after the pulse's maximum. In order to clarify these phenomena, we present spatial and temporal dependencies of the relevant physical quantities with femtosecond and atomic resolution.

\paragraph{Theoretical aspects.} We briefly present the main ideas of our approach to ultrafast electron dynamics, based on our code \textsc{evolve}, which is described in more detail in our Supplemental Material~\cite{Supplement} and elsewhere~\cite{Toepler2021, Ziolkowski23, busch2023a, busch2023b}.

For the purpose of this Paper, we chose a ribbon made of a Cu(001) monolayer. The free-standing film forms a square lattice, with Cartesian axes chosen as $x \equiv [110]$, $y \equiv [\bar{1}10]$, and $z \equiv [001]$ (Fig.~\ref{fig:sketch}). We apply periodic boundary conditions in $x$ direction. The ribbon is $15$~atomic rows wide along $y$.

The electron dynamics is described by the von Neumann equation
\begin{align}
   -\cone \hbar \frac{\mathrm{d} \operator{\rho}(t)}{\mathrm{d} t} & =  [ \operator{\rho}(t),\operator{H}(t)]
    \label{eq:EOM}
\end{align}
for the one-particle density matrix $\operator{\rho}(t)$. The latter is either expressed in a site-orbital basis or in the eigenstate basis of the Hamiltonian $\operator{H}_0$. $\operator{H}_0$ describes the electronic structure of the sample in real space; it is given in tight-binding form and includes spin-orbit coupling~\cite{Slater1954, Papaconstantopoulos2015, Konschuh2010}. 

\begin{figure}
    \centering
    \includegraphics[width=0.99\columnwidth]{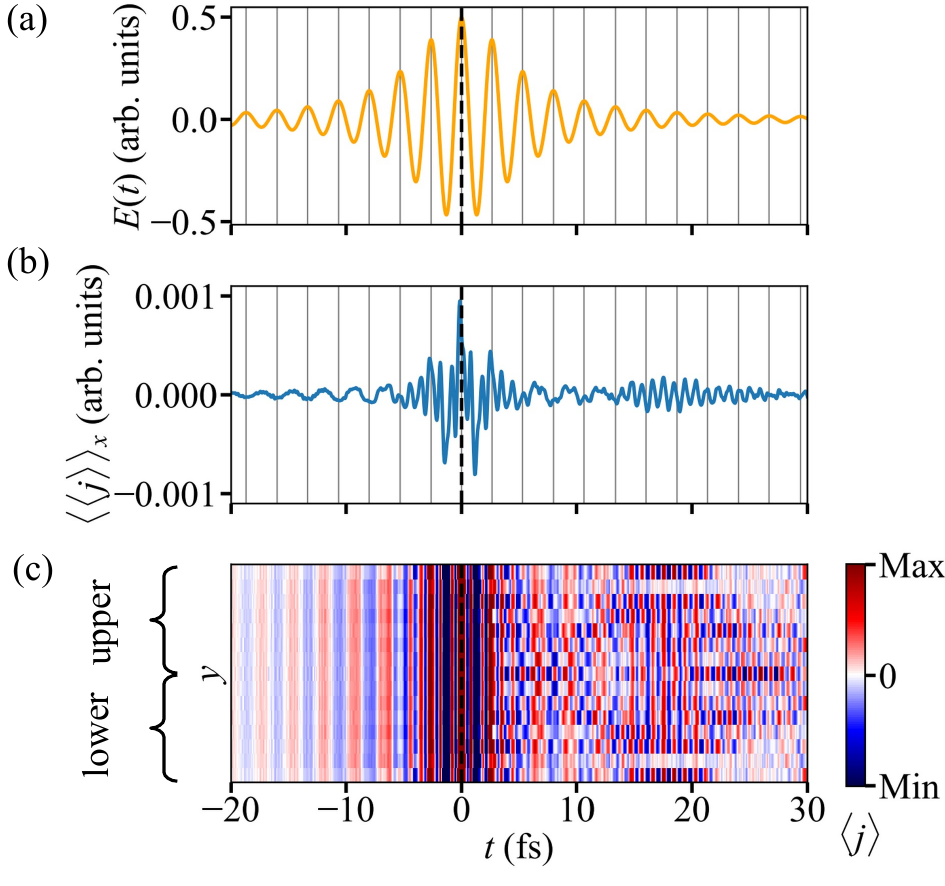}
    \caption{Longitudinal current. (a) Amplitude of the laser pulse with a width of $\unit[10]{fs}$ and centered at $t = \unit[0]{fs}$. (b) Mean longitudinal current~$\expect{\expect{j}}_{x}(t)$; see text. Vertical lines mark maxima of the laser amplitude. (c) Profile of currents of $x$-oriented links (along the ribbon), depicted as color scale. The $y$-average of these currents gives the data shown in panel~b.}
    \label{fig:longitudinal}
\end{figure}

The time-dependent Hamiltonian $\operator{H}(t)$ in~\eqref{eq:EOM} supplements $\operator{H}_0$ by the electric field of the femtosecond laser pulse~\cite{Savasta1995}. This field is a carrier wave of $\hbar \omega = \unit[1.55]{eV}$ energy (equivalent to a period of about $T=\unit[2.7]{fs}$) with a Lorentzian envelope of $\unit[10]{fs}$ width and center at $t = \unit[0]{fs}$ (Fig.~\ref{fig:longitudinal}a); it is linearly polarized and impinges along the $z$-axis (Fig.~\ref{fig:sketch}). The geometry of the entire setup (sample and laser) dictates that only the $z$-component of the OAM $\expect{\vec{L}}$ is produced by the incident radiation~\cite{Henk1996,busch2023a,busch2023b, Supplement}.

Spatio-temporal properties of an observable~$O$ are obtained by taking partial traces in the expectation value $\expect{O}(t) = \operatorname{tr}[\operator{\rho}(t) \,\operator{O}]$, with $\operator{\rho}(t)$ in the site-orbital basis. Besides the occupation probabilities $\expect{p_{k}}(t)$ ($k$ site index), we address currents~\cite{mahan2013,busch2023a}
\begin{align}
	\expect{j_{kl}}(t) & \equiv 
    \frac{\mathrm{i}}{2} \expect{\rho_{lk}(t) \, h_{kl}(t)}
    - \expect{l \leftrightarrow k}
    \label{eq:link-current}
\end{align}
from site $l$ to site $k$. Here, $\rho_{lk}$ and $h_{kl}$ are off-diagonal blocks of the density matrix and of the Hamiltonian matrix in the site representation, respectively. Moreover, we present the $z$-component $\expect{L_{k}^{z}}(t)$ of the OAM at site~$k$ and the $L^{z}$-polarized OAM currents
\begin{align}
	\expect{j_{kl}^{z}}(t) & \equiv \frac{1}{2} \left[ \expect{L^{z} j_{kl}}(t) + \expect{j_{kl} L^{z}}(t) \right]
    \label{eq:link-OAM-current}
\end{align}
from site $l$ to site $k$~\cite{busch2023b}; $L^{z}$ is taken with respect to the sites' positions (atomic center approximation (ACA)~\cite{pezo2022}).\footnote{Focusing on nonequilibrium scenarios, possible equilibrium currents are subtracted. The latter would vanish if angular-momentum currents were properly defined; see Refs.~\onlinecite{sun2005, shi2006} for a discussion of spin currents. Explicit expressions for the currents are given in the Supplemental Material~\cite{Supplement}.} 

Cu exhibits a small spin Hall conductivity, which is attributed to its weak spin-orbit coupling (SOC)~\cite{niimi2011}. In contrast to the spin Hall effect (SHE), the OHE can arise without SOC~\cite{go2018} but requires hybridization of specific atomic orbitals (see e.g.\ Refs.~\onlinecite{ding2020, go2020, go2021, sala2022, pezo2022, busch2023c, canonico2020a, canonico2020b} for more details of the ACA). 
Indeed, all results presented in the following remain qualitatively the same if we neglect SOC thereby confirming the hybridization mechanism as the origin of orbital currents.
In summary, due to its small SOC, Cu is the system of choice in this work since it exhibits a small SHE so that the OHE dominates over the SHE\@.

\paragraph{Laser-induced longitudinal current.} The electric field of the laser drives a current along the nanoribbon (black arrow in Fig.~\ref{fig:sketch}). Averaging currents $\expect{j_{kl}}(t)$ of links $l \to k$ that are oriented in $+x$-direction (Fig.~\ref{fig:longitudinal}c) over the ribbon's width yields the mean longitudinal current $\expect{\expect{j}}_{x}(t)$ (Fig.~\ref{fig:longitudinal}b; for animations of the dynamics, see the Supplemental Material~\cite{Supplement}).

For small laser amplitudes, $\expect{\expect{j}}_{x}(t)$ is expectedly very well correlated with the laser's frequency. However, the time sequence becomes complicated at large field strengths and exhibits a long-period beating pattern (cf.\ the feature at about $\unit[18]{fs}$), which we attribute to various timescales. First, there is the laser period. Second, the electronic structure imposes timescales via the hopping rates in the tight-binding Hamiltonian; these account for the velocities of currents~\cite{Toepler2021}. And third, the inhomogeneous occupation across the ribbon becomes enhanced by the laser-induced dipole transitions, which are stronger the larger the amplitude. The inhomogeneity is clearly seen in the current profile (e.g., at $t > \unit[4]{fs}$ in panel~c).

\begin{figure}
    \centering
    \includegraphics[width = 0.95\columnwidth]{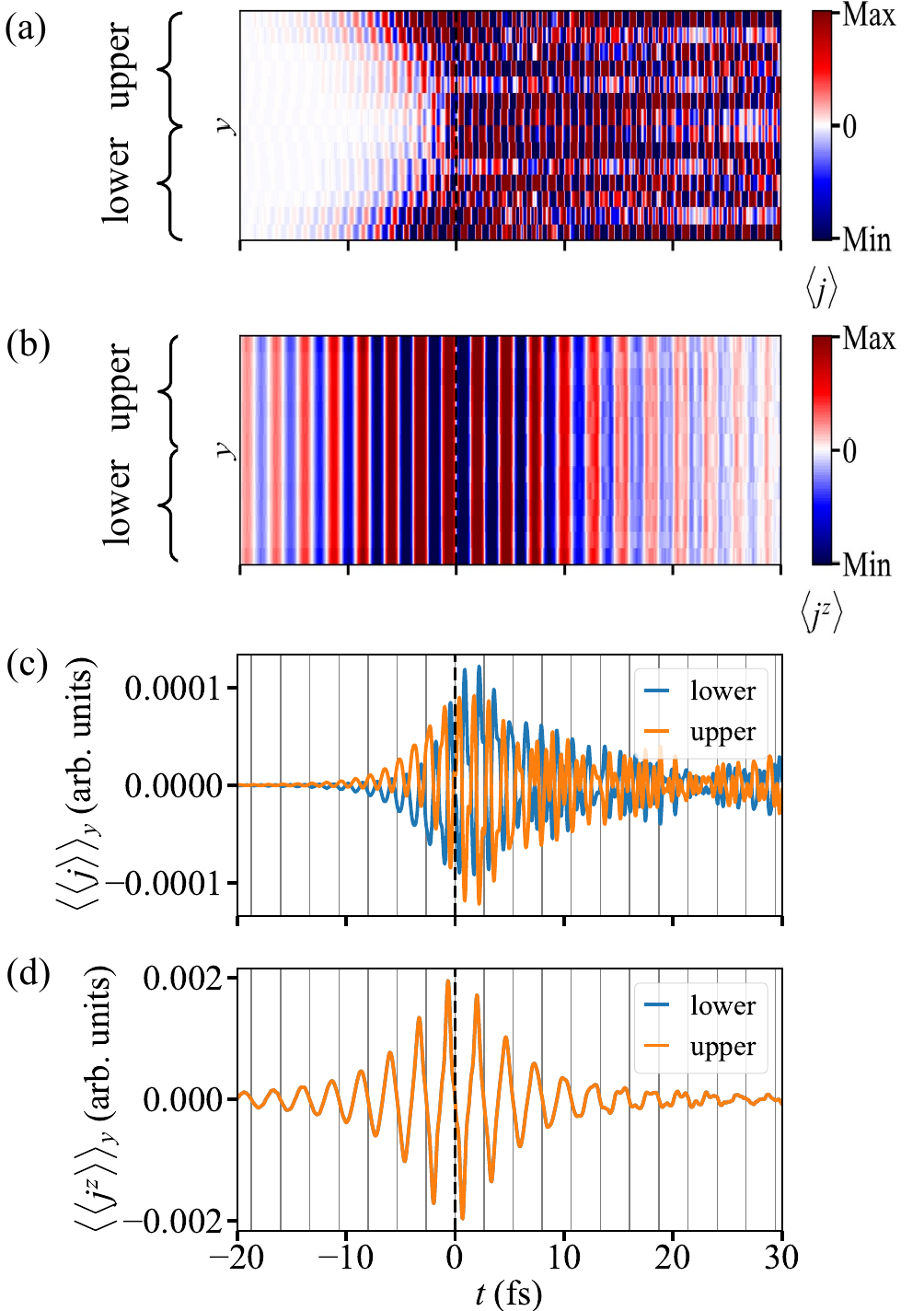}
    \caption{Transversal currents. (a) Profile of transversal link currents depicted as color scale. (b) As panel~a but for the $L^{z}$-polarized OAM currents. (c) and (d) display the data of (a) and (b), respectively, but averaged over the two $y$-regions indicated in panel~a. In panel~d, data for the lower and the upper $y$-region are identical. Vertical lines indicate maxima of the laser's amplitude.}
    \label{fig:transversal}
\end{figure}

\paragraph{Laser-induced transversal currents.}
After the analysis of the longitudinal response (along $x$-direction), we turn now to the transverse response (along $y$-direction). The longitudinal currents are accompanied by transversal currents (along the $y$-direction). Initiated at the edges, an antisymmetric spatio-temporal profile evolves (Fig.~\ref{fig:transversal}a), which develops into a complicated pattern at $t > \unit[0]{fs}$. For example, the link currents in either of the halves (upper/lower regions) do not flow in \emph{one} but in \emph{opposite} directions at a specified time. In other words, the transverse link currents mainly compensate each other in the interior of the ribbon, but not at the edges.  This results in accumulation of charge at the edges as will be demonstrated below. Note that in a closed circuit geometry (without edges) the transverse currents compensate everywhere. 

In order to corroborate the OHE `metaphor' sketched as bent arrows in Fig.~\ref{fig:sketch}, we average the transversal currents over the upper and lower regions. As a result, electrons flow -- on average -- from the center toward the ribbon edges or vice versa (panel~c). Moreover, they oscillate with half a period of the laser's (e.g., at $t < \unit[0]{fs}$): regardless of the orientation of the electric field, electrons moving along the ribbon are `deflected' toward the edges. In one half-cycle of the laser pulse, however, positive OAM is transported toward the upper edge (Fig.~\ref{fig:sketch}a), in the next half-cycle toward the lower edge (Fig.~\ref{fig:sketch}b): the transverse OAM current follows the period of the laser pulse, as seen in Fig.~\ref{fig:transversal}d.

\begin{figure}
    \centering
    \includegraphics[width=0.9\columnwidth]{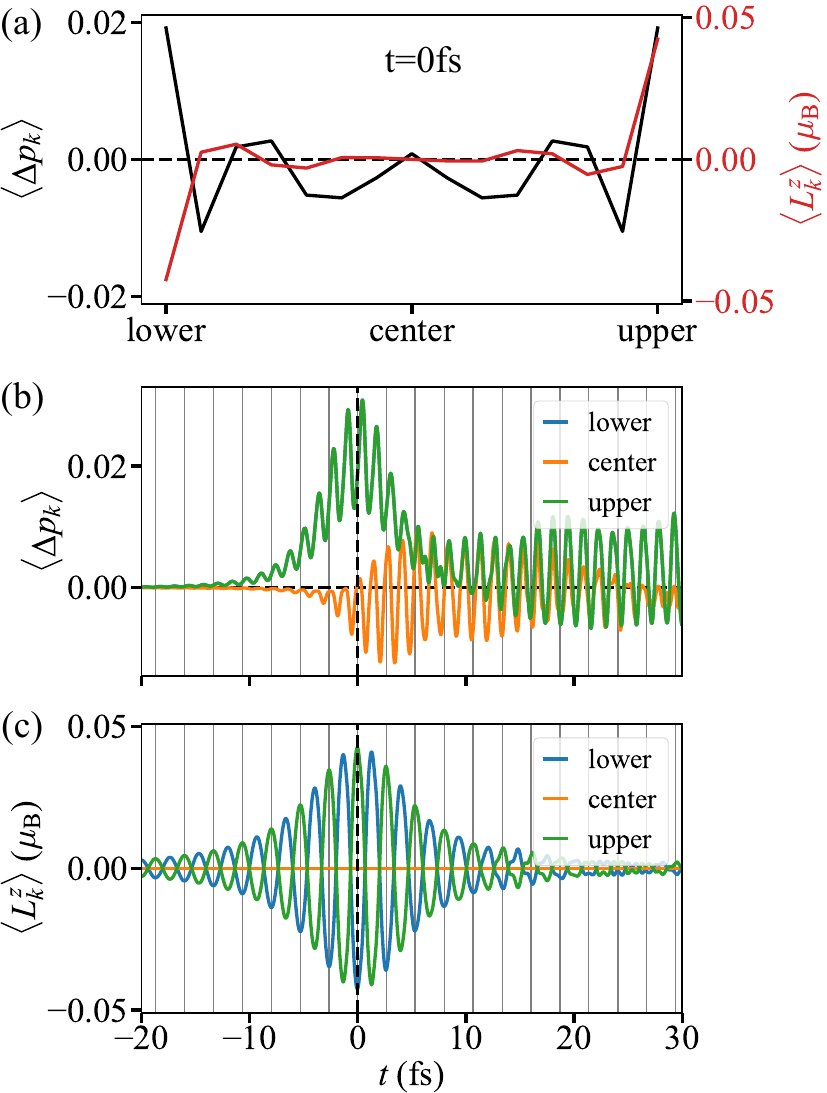}
    \caption{Accumulation of occupation and orbital angular momentum. (a) Occupation profile $\expect{\Delta p_{k}}$ across the ribbon (i.e., in $y$-direction) at $t = \unit[0]{fs}$ relative to the equilibrium profile (black; left abscissa). The respective OAM profile $\expect{L_{k}^{z}}$ is shown in red (right abscissa). (b) As panel~a, but $\expect{\Delta p_{k}}$ versus time~$t$ for two edge sites (blue: lower edge; green: upper edge) and a central site (orange). Both edges have identical occupation (cf.\ panel~a). (c) As panel~b but for $\expect{L_{k}^{z}}$. Opposite edges exhibit OAM with opposite signs. Vertical lines mark the maxima of the laser's amplitude.}
    \label{fig:accumulation}
\end{figure}

Besides its period, the region-averaged transversal current differs from its OAM companion within more aspects. The OAM current appears much more `regular', in particular it is homogeneous across the sample and exhibits a clear-cut time signature, while the link current does not (compare panel~a with panel~b). It is sizable only during the pulse, in contrast to the current which `persists' after the pulse (compare panel~c with panel~d at $t \approx \unit[25]{fs}$). They have in common that their extrema do not coincide with those of the laser amplitude; there is a time difference of about $T/4$. % $\unit[1]{fs}$.

We attribute the mentioned differences to the fact that the link current comprises a flow of $L^{z}$-polarized \emph{and} -unpolarized electrons, while the $L^{z}$-current accounts only for polarized electrons. Following up on the above reasoning concerning different timescales that are related to the hopping rates in the Hamiltonian, this astriction to one class of orbitals gives a clearer time signature of the OAM current. 

\paragraph{Laser-induced OAM and occupation.} In order to discuss accumulation of OAM at the ribbon's edges, we present occupation $\expect{\Delta p_{k}}$ (i.e. the change of occupation with respect to equilibrium) and OAM $\expect{L^{z}_{k}}$ profiles at $t = \unit[0]{fs}$ (Fig.~\ref{fig:accumulation}a). Both quantities are largest (in absolute value) at edge sites and exhibit significantly smaller values in the sample's interior. In particular $L^{z}$ (red curve) is strongly localized at edge sites, exhibiting an absolute value as large as about $\unit[0.05]{\mu_{\mathrm{B}}}$.

In accordance with the antisymmetry of the transversal-current profile (Figs.~\ref{fig:transversal}a and~c), both edges are identically populated, which corresponds to charge accumulation at the edges, (black in Fig.~\ref{fig:accumulation}a; there is no Hall voltage) and show half the laser's oscillation period (panel~b) since electrons with opposite $\expect{L^z}$ accumulate comparing two neighboring maxima. Closer inspection shows a small time lag with respect to the laser amplitude. More specifically, the extrema of the occupation of the edge sites and a central site are slightly out of phase, which is explained by the finite velocity (determined by the hoppings) of the transversal flow. 

As has been briefly addressed, the edge occupation comprises $L^{z}$-unpolarized and -polarized contributions. Only the latter lead to accumulation of OAM (red in panel~a). The OAM time series exhibits opposite signs (panel~c), in line with the oppositely $L^{z}$-polarized transversal flow (Fig.~\ref{fig:transversal}b), a finding supporting the oppositely oriented $L^{z}$-arrows in Fig.~\ref{fig:sketch}.

The ultrafast OHE cannot be understood as a time series of steady orbital Hall effects for varying electric field. In a closed circuit geometry, where no accumulation of OAM is possible, we still observe the same phase shifts between laser and orbital current in contrast to a steady orbital Hall effect where $\vec{j}_{L}$ is proportional to $\vec{E}$ (cf. Fig.~\ref{fig:sketch}). Instead, it is laser driven: A maximum orbital current is present when the $E$ field is zero akin to a driven harmonic oscillator which exhibits the same phase relations between driving force and velocity
(cf.\  Figs.~\ref{fig:longitudinal}a, \ref{fig:transversal}d, and the Supplemental Material \cite{Supplement}). 

\paragraph{Concluding remarks.} Our detailed theoretical analysis establishes an extension of the well-known static orbital Hall effect onto ultrashort timescales. On the one hand, it fully supports the pictorial presentation of the OHE (Fig.~\ref{fig:sketch}). On the other hand, we point out differences between OAM-carrying and non-OAM-carrying quantities, such as OAM currents and (charge) currents. Additionally, the UOHE exhibits specific phase relations between accumulated OAM, OAM currents, and the electric field of the laser. This finding strongly suggests that the UOHE is actually laser-driven and not a chronology of steady orbital Hall effects, with each `snapshot' taken for the respective electric field (see Ref.~\onlinecite{Supplement}). Furthermore, since the OHE has been analyzed with an oscillating field, we can observe the accumulation of both OAM and OAM currents. In a static OHE, only one quantity can be analyzed, depending on the considered boundary conditions.

Our results for Cu nanoribbons call for studies of other materials. 
%Graphene is a promising candidate considering its technological importance~\cite{akinwande2019,wang2020}. 
As has been shown in Ref.~\onlinecite{salemi2023} the OHE is larger than the SHE in most metals.
% Why graphene? Maybe you can also mention the elements where OHE is large and SHE is small in this nice paper:
% PRMaterials 6, 095001 (2022) Fig. 1
% This is mainly Sc, Ti, V, Cr, Mn, ...
% But is may be a bit risky because the question may arise why we investigated Cu that has a comparably low OHE/SHE ratio compared to these materials.  
It is also worth investigating ferromagnets, where a comparison of ultrafast anomalous, spin, and orbital Hall effects comes to mind. In a recent experiment on ferromagnet/normal metal heterostructures excited by femtosecond laser pulses, THz emission spectroscopy was utilized to identify OAM currents~\cite{seifert2023}.

This work is funded by the Deutsche Forschungsgemeinschaft (DFG, German Research Foundation) -- Project-ID 328545488 -- TRR~227, project~B04.

\bibliographystyle{apsrev4-2}
\bibliography{references}

\begin{thebibliography}{62}%
\makeatletter
\providecommand \@ifxundefined [1]{%
 \@ifx{#1\undefined}
}%
\providecommand \@ifnum [1]{%
 \ifnum #1\expandafter \@firstoftwo
 \else \expandafter \@secondoftwo
 \fi
}%
\providecommand \@ifx [1]{%
 \ifx #1\expandafter \@firstoftwo
 \else \expandafter \@secondoftwo
 \fi
}%
\providecommand \natexlab [1]{#1}%
\providecommand \enquote  [1]{``#1''}%
\providecommand \bibnamefont  [1]{#1}%
\providecommand \bibfnamefont [1]{#1}%
\providecommand \citenamefont [1]{#1}%
\providecommand \href@noop [0]{\@secondoftwo}%
\providecommand \href [0]{\begingroup \@sanitize@url \@href}%
\providecommand \@href[1]{\@@startlink{#1}\@@href}%
\providecommand \@@href[1]{\endgroup#1\@@endlink}%
\providecommand \@sanitize@url [0]{\catcode `\\12\catcode `\$12\catcode
  `\&12\catcode `\#12\catcode `\^12\catcode `\_12\catcode `\%12\relax}%
\providecommand \@@startlink[1]{}%
\providecommand \@@endlink[0]{}%
\providecommand \url  [0]{\begingroup\@sanitize@url \@url }%
\providecommand \@url [1]{\endgroup\@href {#1}{\urlprefix }}%
\providecommand \urlprefix  [0]{URL }%
\providecommand \Eprint [0]{\href }%
\providecommand \doibase [0]{https://doi.org/}%
\providecommand \selectlanguage [0]{\@gobble}%
\providecommand \bibinfo  [0]{\@secondoftwo}%
\providecommand \bibfield  [0]{\@secondoftwo}%
\providecommand \translation [1]{[#1]}%
\providecommand \BibitemOpen [0]{}%
\providecommand \bibitemStop [0]{}%
\providecommand \bibitemNoStop [0]{.\EOS\space}%
\providecommand \EOS [0]{\spacefactor3000\relax}%
\providecommand \BibitemShut  [1]{\csname bibitem#1\endcsname}%
\let\auto@bib@innerbib\@empty
%</preamble>
\bibitem [{\citenamefont {Cao}\ \emph {et~al.}(2020)\citenamefont {Cao},
  \citenamefont {Xing}, \citenamefont {Lin}, \citenamefont {Zhang},
  \citenamefont {Zheng},\ and\ \citenamefont {Wang}}]{cao2020}%
  \BibitemOpen
  \bibfield  {author} {\bibinfo {author} {\bibfnamefont {Y.}~\bibnamefont
  {Cao}}, \bibinfo {author} {\bibfnamefont {G.}~\bibnamefont {Xing}}, \bibinfo
  {author} {\bibfnamefont {H.}~\bibnamefont {Lin}}, \bibinfo {author}
  {\bibfnamefont {N.}~\bibnamefont {Zhang}}, \bibinfo {author} {\bibfnamefont
  {H.}~\bibnamefont {Zheng}},\ and\ \bibinfo {author} {\bibfnamefont
  {K.}~\bibnamefont {Wang}},\ }\href
  {https://doi.org/https://doi.org/10.1016/j.isci.2020.101614} {\bibfield
  {journal} {\bibinfo  {journal} {iScience}\ }\textbf {\bibinfo {volume}
  {23}},\ \bibinfo {pages} {101614} (\bibinfo {year} {2020})}\BibitemShut
  {NoStop}%
\bibitem [{\citenamefont {Go}\ \emph {et~al.}(2021)\citenamefont {Go},
  \citenamefont {Jo}, \citenamefont {Lee}, \citenamefont {Kl{\"a}ui},\ and\
  \citenamefont {Mokrousov}}]{go2021}%
  \BibitemOpen
  \bibfield  {author} {\bibinfo {author} {\bibfnamefont {D.}~\bibnamefont
  {Go}}, \bibinfo {author} {\bibfnamefont {D.}~\bibnamefont {Jo}}, \bibinfo
  {author} {\bibfnamefont {H.-W.}\ \bibnamefont {Lee}}, \bibinfo {author}
  {\bibfnamefont {M.}~\bibnamefont {Kl{\"a}ui}},\ and\ \bibinfo {author}
  {\bibfnamefont {Y.}~\bibnamefont {Mokrousov}},\ }\href
  {https://doi.org/10.1209/0295-5075/ac2653} {\bibfield  {journal} {\bibinfo
  {journal} {Europhysics Letters}\ }\textbf {\bibinfo {volume} {135}},\
  \bibinfo {pages} {37001} (\bibinfo {year} {2021})}\BibitemShut {NoStop}%
\bibitem [{\citenamefont {Rappoport}(2023)}]{rappoport2023}%
  \BibitemOpen
  \bibfield  {author} {\bibinfo {author} {\bibfnamefont {T.~G.}\ \bibnamefont
  {Rappoport}},\ }\href
  {https://doi.org/https://doi.org/10.1038/d41586-023-02072-z} {\bibinfo
  {title} {First light on orbitronics as a viable alternative to electronics}}
  (\bibinfo {year} {2023})\BibitemShut {NoStop}%
\bibitem [{\citenamefont {D'yakonov}\ and\ \citenamefont
  {Perel}(1971)}]{dyakonov1971}%
  \BibitemOpen
  \bibfield  {author} {\bibinfo {author} {\bibfnamefont {M.~I.}\ \bibnamefont
  {D'yakonov}}\ and\ \bibinfo {author} {\bibfnamefont {V.~I.}\ \bibnamefont
  {Perel}},\ }\href
  {https://doi.org/https://doi.org/10.1016/0375-9601(71)90196-4} {\bibfield
  {journal} {\bibinfo  {journal} {Physics Letters A}\ }\textbf {\bibinfo
  {volume} {35}},\ \bibinfo {pages} {459} (\bibinfo {year} {1971})}\BibitemShut
  {NoStop}%
\bibitem [{\citenamefont {Hirsch}(1999)}]{hirsch1999}%
  \BibitemOpen
  \bibfield  {author} {\bibinfo {author} {\bibfnamefont {J.~E.}\ \bibnamefont
  {Hirsch}},\ }\href {https://doi.org/10.1103/PhysRevLett.83.1834} {\bibfield
  {journal} {\bibinfo  {journal} {Physical Review Letters}\ }\textbf {\bibinfo
  {volume} {83}},\ \bibinfo {pages} {1834} (\bibinfo {year}
  {1999})}\BibitemShut {NoStop}%
\bibitem [{\citenamefont {Kato}\ \emph {et~al.}(2004)\citenamefont {Kato},
  \citenamefont {Myers}, \citenamefont {Gossard},\ and\ \citenamefont
  {Awschalom}}]{kato2004}%
  \BibitemOpen
  \bibfield  {author} {\bibinfo {author} {\bibfnamefont {Y.~K.}\ \bibnamefont
  {Kato}}, \bibinfo {author} {\bibfnamefont {R.~C.}\ \bibnamefont {Myers}},
  \bibinfo {author} {\bibfnamefont {A.~C.}\ \bibnamefont {Gossard}},\ and\
  \bibinfo {author} {\bibfnamefont {D.~D.}\ \bibnamefont {Awschalom}},\ }\href
  {https://doi.org/10.1126/science.1105514} {\bibfield  {journal} {\bibinfo
  {journal} {Science}\ }\textbf {\bibinfo {volume} {306}},\ \bibinfo {pages}
  {1910} (\bibinfo {year} {2004})}\BibitemShut {NoStop}%
\bibitem [{\citenamefont {Sinova}\ \emph {et~al.}(2015)\citenamefont {Sinova},
  \citenamefont {Valenzuela}, \citenamefont {Wunderlich}, \citenamefont
  {Back},\ and\ \citenamefont {Jungwirth}}]{sinova2015}%
  \BibitemOpen
  \bibfield  {author} {\bibinfo {author} {\bibfnamefont {J.}~\bibnamefont
  {Sinova}}, \bibinfo {author} {\bibfnamefont {S.~O.}\ \bibnamefont
  {Valenzuela}}, \bibinfo {author} {\bibfnamefont {J.}~\bibnamefont
  {Wunderlich}}, \bibinfo {author} {\bibfnamefont {C.}~\bibnamefont {Back}},\
  and\ \bibinfo {author} {\bibfnamefont {T.}~\bibnamefont {Jungwirth}},\ }\href
  {https://doi.org/10.1103/RevModPhys.87.1213} {\bibfield  {journal} {\bibinfo
  {journal} {Reviews of Modern Physics}\ }\textbf {\bibinfo {volume} {87}},\
  \bibinfo {pages} {1213} (\bibinfo {year} {2015})}\BibitemShut {NoStop}%
\bibitem [{\citenamefont {{Aronov}}\ and\ \citenamefont
  {{Lyanda-Geller}}(1989)}]{Aronov1989}%
  \BibitemOpen
  \bibfield  {author} {\bibinfo {author} {\bibfnamefont {A.~G.}\ \bibnamefont
  {{Aronov}}}\ and\ \bibinfo {author} {\bibfnamefont {Y.~B.}\ \bibnamefont
  {{Lyanda-Geller}}},\ }\href@noop {} {\bibfield  {journal} {\bibinfo
  {journal} {Soviet Journal of Experimental and Theoretical Physics Letters}\
  }\textbf {\bibinfo {volume} {50}},\ \bibinfo {pages} {431} (\bibinfo {year}
  {1989})}\BibitemShut {NoStop}%
\bibitem [{\citenamefont {Edelstein}(1990)}]{edelstein1990}%
  \BibitemOpen
  \bibfield  {author} {\bibinfo {author} {\bibfnamefont {V.}~\bibnamefont
  {Edelstein}},\ }\href
  {https://doi.org/https://doi.org/10.1016/0038-1098(90)90963-C} {\bibfield
  {journal} {\bibinfo  {journal} {Solid State Communications}\ }\textbf
  {\bibinfo {volume} {73}},\ \bibinfo {pages} {233} (\bibinfo {year}
  {1990})}\BibitemShut {NoStop}%
\bibitem [{\citenamefont {Inoue}\ \emph {et~al.}(2003)\citenamefont {Inoue},
  \citenamefont {Bauer},\ and\ \citenamefont {Molenkamp}}]{inoue2003}%
  \BibitemOpen
  \bibfield  {author} {\bibinfo {author} {\bibfnamefont {J.-i.}\ \bibnamefont
  {Inoue}}, \bibinfo {author} {\bibfnamefont {G.~E.~W.}\ \bibnamefont
  {Bauer}},\ and\ \bibinfo {author} {\bibfnamefont {L.~W.}\ \bibnamefont
  {Molenkamp}},\ }\href {https://doi.org/10.1103/PhysRevB.67.033104} {\bibfield
   {journal} {\bibinfo  {journal} {Physical Review B}\ }\textbf {\bibinfo
  {volume} {67}},\ \bibinfo {pages} {033104} (\bibinfo {year}
  {2003})}\BibitemShut {NoStop}%
\bibitem [{\citenamefont {Gambardella}\ and\ \citenamefont
  {Miron}(2011)}]{gambardella2011}%
  \BibitemOpen
  \bibfield  {author} {\bibinfo {author} {\bibfnamefont {P.}~\bibnamefont
  {Gambardella}}\ and\ \bibinfo {author} {\bibfnamefont {I.~M.}\ \bibnamefont
  {Miron}},\ }\href {https://doi.org/https://doi.org/10.1098/rsta.2010.0336}
  {\bibfield  {journal} {\bibinfo  {journal} {Philosophical Transactions of the
  Royal Society A}\ }\textbf {\bibinfo {volume} {369}},\ \bibinfo {pages}
  {3175–3197} (\bibinfo {year} {2011})}\BibitemShut {NoStop}%
\bibitem [{\citenamefont {Zhang}\ and\ \citenamefont {Yang}(2005)}]{zhang2005}%
  \BibitemOpen
  \bibfield  {author} {\bibinfo {author} {\bibfnamefont {S.}~\bibnamefont
  {Zhang}}\ and\ \bibinfo {author} {\bibfnamefont {Z.}~\bibnamefont {Yang}},\
  }\href {https://doi.org/10.1103/PhysRevLett.94.066602} {\bibfield  {journal}
  {\bibinfo  {journal} {Physical Review Letters}\ }\textbf {\bibinfo {volume}
  {94}},\ \bibinfo {pages} {066602} (\bibinfo {year} {2005})}\BibitemShut
  {NoStop}%
\bibitem [{\citenamefont {Bernevig}\ \emph {et~al.}(2005)\citenamefont
  {Bernevig}, \citenamefont {Hughes},\ and\ \citenamefont
  {Zhang}}]{bernevig2005}%
  \BibitemOpen
  \bibfield  {author} {\bibinfo {author} {\bibfnamefont {B.~A.}\ \bibnamefont
  {Bernevig}}, \bibinfo {author} {\bibfnamefont {T.~L.}\ \bibnamefont
  {Hughes}},\ and\ \bibinfo {author} {\bibfnamefont {S.-C.}\ \bibnamefont
  {Zhang}},\ }\href {https://doi.org/10.1103/PhysRevLett.95.066601} {\bibfield
  {journal} {\bibinfo  {journal} {Physical Review Letters}\ }\textbf {\bibinfo
  {volume} {95}},\ \bibinfo {pages} {066601} (\bibinfo {year}
  {2005})}\BibitemShut {NoStop}%
\bibitem [{\citenamefont {Kontani}\ \emph {et~al.}(2008)\citenamefont
  {Kontani}, \citenamefont {Tanaka}, \citenamefont {Hirashima}, \citenamefont
  {Yamada},\ and\ \citenamefont {Inoue}}]{kontani2008}%
  \BibitemOpen
  \bibfield  {author} {\bibinfo {author} {\bibfnamefont {H.}~\bibnamefont
  {Kontani}}, \bibinfo {author} {\bibfnamefont {T.}~\bibnamefont {Tanaka}},
  \bibinfo {author} {\bibfnamefont {D.}~\bibnamefont {Hirashima}}, \bibinfo
  {author} {\bibfnamefont {K.}~\bibnamefont {Yamada}},\ and\ \bibinfo {author}
  {\bibfnamefont {J.}~\bibnamefont {Inoue}},\ }\href
  {https://doi.org/10.1103/PhysRevLett.100.096601} {\bibfield  {journal}
  {\bibinfo  {journal} {Physical Review Letters}\ }\textbf {\bibinfo {volume}
  {100}},\ \bibinfo {pages} {096601} (\bibinfo {year} {2008})}\BibitemShut
  {NoStop}%
\bibitem [{\citenamefont {Tanaka}\ \emph {et~al.}(2008)\citenamefont {Tanaka},
  \citenamefont {Kontani}, \citenamefont {Naito}, \citenamefont {Naito},
  \citenamefont {Hirashima}, \citenamefont {Yamada},\ and\ \citenamefont
  {Inoue}}]{tanaka2008}%
  \BibitemOpen
  \bibfield  {author} {\bibinfo {author} {\bibfnamefont {T.}~\bibnamefont
  {Tanaka}}, \bibinfo {author} {\bibfnamefont {H.}~\bibnamefont {Kontani}},
  \bibinfo {author} {\bibfnamefont {M.}~\bibnamefont {Naito}}, \bibinfo
  {author} {\bibfnamefont {T.}~\bibnamefont {Naito}}, \bibinfo {author}
  {\bibfnamefont {D.~S.}\ \bibnamefont {Hirashima}}, \bibinfo {author}
  {\bibfnamefont {K.}~\bibnamefont {Yamada}},\ and\ \bibinfo {author}
  {\bibfnamefont {J.}~\bibnamefont {Inoue}},\ }\href
  {https://doi.org/10.1103/PhysRevB.77.165117} {\bibfield  {journal} {\bibinfo
  {journal} {Physical Review B}\ }\textbf {\bibinfo {volume} {77}},\ \bibinfo
  {pages} {165117} (\bibinfo {year} {2008})}\BibitemShut {NoStop}%
\bibitem [{\citenamefont {Kontani}\ \emph {et~al.}(2009)\citenamefont
  {Kontani}, \citenamefont {Tanaka}, \citenamefont {Hirashima}, \citenamefont
  {Yamada},\ and\ \citenamefont {Inoue}}]{kontani2009}%
  \BibitemOpen
  \bibfield  {author} {\bibinfo {author} {\bibfnamefont {H.}~\bibnamefont
  {Kontani}}, \bibinfo {author} {\bibfnamefont {T.}~\bibnamefont {Tanaka}},
  \bibinfo {author} {\bibfnamefont {D.}~\bibnamefont {Hirashima}}, \bibinfo
  {author} {\bibfnamefont {K.}~\bibnamefont {Yamada}},\ and\ \bibinfo {author}
  {\bibfnamefont {J.}~\bibnamefont {Inoue}},\ }\href
  {https://doi.org/10.1103/PhysRevLett.102.016601} {\bibfield  {journal}
  {\bibinfo  {journal} {Physical Review Letters}\ }\textbf {\bibinfo {volume}
  {102}},\ \bibinfo {pages} {016601} (\bibinfo {year} {2009})}\BibitemShut
  {NoStop}%
\bibitem [{\citenamefont {Zhong}\ \emph {et~al.}(2016)\citenamefont {Zhong},
  \citenamefont {Moore},\ and\ \citenamefont {Souza}}]{zhong2016}%
  \BibitemOpen
  \bibfield  {author} {\bibinfo {author} {\bibfnamefont {S.}~\bibnamefont
  {Zhong}}, \bibinfo {author} {\bibfnamefont {J.~E.}\ \bibnamefont {Moore}},\
  and\ \bibinfo {author} {\bibfnamefont {I.}~\bibnamefont {Souza}},\ }\href
  {https://doi.org/10.1103/PhysRevLett.116.077201} {\bibfield  {journal}
  {\bibinfo  {journal} {Physical Review Letters}\ }\textbf {\bibinfo {volume}
  {116}},\ \bibinfo {pages} {077201} (\bibinfo {year} {2016})}\BibitemShut
  {NoStop}%
\bibitem [{\citenamefont {Yoda}\ \emph {et~al.}(2018)\citenamefont {Yoda},
  \citenamefont {Yokoyama},\ and\ \citenamefont {Murakami}}]{yoda2018}%
  \BibitemOpen
  \bibfield  {author} {\bibinfo {author} {\bibfnamefont {T.}~\bibnamefont
  {Yoda}}, \bibinfo {author} {\bibfnamefont {T.}~\bibnamefont {Yokoyama}},\
  and\ \bibinfo {author} {\bibfnamefont {S.}~\bibnamefont {Murakami}},\ }\href
  {https://doi.org/10.1021/acs.nanolett.7b04300} {\bibfield  {journal}
  {\bibinfo  {journal} {Nano Letters}\ }\textbf {\bibinfo {volume} {18}},\
  \bibinfo {pages} {916} (\bibinfo {year} {2018})}\BibitemShut {NoStop}%
\bibitem [{\citenamefont {Salemi}\ \emph {et~al.}(2019)\citenamefont {Salemi},
  \citenamefont {Berritta}, \citenamefont {Nandy},\ and\ \citenamefont
  {Oppeneer}}]{salemi2019}%
  \BibitemOpen
  \bibfield  {author} {\bibinfo {author} {\bibfnamefont {L.}~\bibnamefont
  {Salemi}}, \bibinfo {author} {\bibfnamefont {M.}~\bibnamefont {Berritta}},
  \bibinfo {author} {\bibfnamefont {A.~K.}\ \bibnamefont {Nandy}},\ and\
  \bibinfo {author} {\bibfnamefont {P.~M.}\ \bibnamefont {Oppeneer}},\ }\href
  {https://doi.org/10.1038/s41467-019-13367-z} {\bibfield  {journal} {\bibinfo
  {journal} {Nature Communications}\ }\textbf {\bibinfo {volume} {10}},\
  \bibinfo {pages} {5381} (\bibinfo {year} {2019})}\BibitemShut {NoStop}%
\bibitem [{\citenamefont {Johansson}\ \emph {et~al.}(2021)\citenamefont
  {Johansson}, \citenamefont {G\"obel}, \citenamefont {Henk}, \citenamefont
  {Bibes},\ and\ \citenamefont {Mertig}}]{johansson2021}%
  \BibitemOpen
  \bibfield  {author} {\bibinfo {author} {\bibfnamefont {A.}~\bibnamefont
  {Johansson}}, \bibinfo {author} {\bibfnamefont {B.}~\bibnamefont {G\"obel}},
  \bibinfo {author} {\bibfnamefont {J.}~\bibnamefont {Henk}}, \bibinfo {author}
  {\bibfnamefont {M.}~\bibnamefont {Bibes}},\ and\ \bibinfo {author}
  {\bibfnamefont {I.}~\bibnamefont {Mertig}},\ }\href
  {https://doi.org/10.1103/PhysRevResearch.3.013275} {\bibfield  {journal}
  {\bibinfo  {journal} {Physical Review Research}\ }\textbf {\bibinfo {volume}
  {3}},\ \bibinfo {pages} {013275} (\bibinfo {year} {2021})}\BibitemShut
  {NoStop}%
\bibitem [{\citenamefont {Hafez}\ \emph {et~al.}(2016)\citenamefont {Hafez},
  \citenamefont {Chai}, \citenamefont {Ibrahim}, \citenamefont {Mondal},
  \citenamefont {Férachou}, \citenamefont {Ropagnol},\ and\ \citenamefont
  {Ozaki}}]{hafez2016}%
  \BibitemOpen
  \bibfield  {author} {\bibinfo {author} {\bibfnamefont {H.~A.}\ \bibnamefont
  {Hafez}}, \bibinfo {author} {\bibfnamefont {X.}~\bibnamefont {Chai}},
  \bibinfo {author} {\bibfnamefont {A.}~\bibnamefont {Ibrahim}}, \bibinfo
  {author} {\bibfnamefont {S.}~\bibnamefont {Mondal}}, \bibinfo {author}
  {\bibfnamefont {D.}~\bibnamefont {Férachou}}, \bibinfo {author}
  {\bibfnamefont {X.}~\bibnamefont {Ropagnol}},\ and\ \bibinfo {author}
  {\bibfnamefont {T.}~\bibnamefont {Ozaki}},\ }\href
  {https://doi.org/10.1088/2040-8978/18/9/093004} {\bibfield  {journal}
  {\bibinfo  {journal} {Journal of Optics}\ }\textbf {\bibinfo {volume} {18}},\
  \bibinfo {pages} {093004} (\bibinfo {year} {2016})}\BibitemShut {NoStop}%
\bibitem [{\citenamefont {Phillips}\ \emph {et~al.}(2015)\citenamefont
  {Phillips}, \citenamefont {Gandhi}, \citenamefont {Mazur},\ and\
  \citenamefont {Sundaram}}]{phillips2015}%
  \BibitemOpen
  \bibfield  {author} {\bibinfo {author} {\bibfnamefont {K.~C.}\ \bibnamefont
  {Phillips}}, \bibinfo {author} {\bibfnamefont {H.~H.}\ \bibnamefont
  {Gandhi}}, \bibinfo {author} {\bibfnamefont {E.}~\bibnamefont {Mazur}},\ and\
  \bibinfo {author} {\bibfnamefont {S.~K.}\ \bibnamefont {Sundaram}},\ }\href
  {https://doi.org/10.1364/AOP.7.000684} {\bibfield  {journal} {\bibinfo
  {journal} {Advances in Optics and Photonics}\ }\textbf {\bibinfo {volume}
  {7}},\ \bibinfo {pages} {684} (\bibinfo {year} {2015})}\BibitemShut {NoStop}%
\bibitem [{\citenamefont {Seemann}\ \emph {et~al.}(2015)\citenamefont
  {Seemann}, \citenamefont {K\"odderitzsch}, \citenamefont {Wimmer},\ and\
  \citenamefont {Ebert}}]{seemann2015}%
  \BibitemOpen
  \bibfield  {author} {\bibinfo {author} {\bibfnamefont {M.}~\bibnamefont
  {Seemann}}, \bibinfo {author} {\bibfnamefont {D.}~\bibnamefont
  {K\"odderitzsch}}, \bibinfo {author} {\bibfnamefont {S.}~\bibnamefont
  {Wimmer}},\ and\ \bibinfo {author} {\bibfnamefont {H.}~\bibnamefont
  {Ebert}},\ }\href {https://doi.org/10.1103/PhysRevB.92.155138} {\bibfield
  {journal} {\bibinfo  {journal} {Physical Review B}\ }\textbf {\bibinfo
  {volume} {92}},\ \bibinfo {pages} {155138} (\bibinfo {year}
  {2015})}\BibitemShut {NoStop}%
\bibitem [{\citenamefont {Nagaosa}\ \emph {et~al.}(2010)\citenamefont
  {Nagaosa}, \citenamefont {Sinova}, \citenamefont {Onoda}, \citenamefont
  {MacDonald},\ and\ \citenamefont {Ong}}]{nagaosa2010}%
  \BibitemOpen
  \bibfield  {author} {\bibinfo {author} {\bibfnamefont {N.}~\bibnamefont
  {Nagaosa}}, \bibinfo {author} {\bibfnamefont {J.}~\bibnamefont {Sinova}},
  \bibinfo {author} {\bibfnamefont {S.}~\bibnamefont {Onoda}}, \bibinfo
  {author} {\bibfnamefont {A.~H.}\ \bibnamefont {MacDonald}},\ and\ \bibinfo
  {author} {\bibfnamefont {N.~P.}\ \bibnamefont {Ong}},\ }\href
  {https://doi.org/10.1103/RevModPhys.82.1539} {\bibfield  {journal} {\bibinfo
  {journal} {Reviews of Modern Physics}\ }\textbf {\bibinfo {volume} {82}},\
  \bibinfo {pages} {1539} (\bibinfo {year} {2010})}\BibitemShut {NoStop}%
\bibitem [{\citenamefont {Go}\ \emph {et~al.}(2018)\citenamefont {Go},
  \citenamefont {Jo}, \citenamefont {Kim},\ and\ \citenamefont {Lee}}]{go2018}%
  \BibitemOpen
  \bibfield  {author} {\bibinfo {author} {\bibfnamefont {D.}~\bibnamefont
  {Go}}, \bibinfo {author} {\bibfnamefont {D.}~\bibnamefont {Jo}}, \bibinfo
  {author} {\bibfnamefont {C.}~\bibnamefont {Kim}},\ and\ \bibinfo {author}
  {\bibfnamefont {H.-W.}\ \bibnamefont {Lee}},\ }\href
  {https://doi.org/10.1103/PhysRevLett.121.086602} {\bibfield  {journal}
  {\bibinfo  {journal} {Physical Review Letters}\ }\textbf {\bibinfo {volume}
  {121}},\ \bibinfo {pages} {086602} (\bibinfo {year} {2018})}\BibitemShut
  {NoStop}%
\bibitem [{\citenamefont {de~Juan}\ \emph {et~al.}(2017)\citenamefont
  {de~Juan}, \citenamefont {Grushin}, \citenamefont {Morimoto},\ and\
  \citenamefont {Moore}}]{dejuan2017}%
  \BibitemOpen
  \bibfield  {author} {\bibinfo {author} {\bibfnamefont {F.}~\bibnamefont
  {de~Juan}}, \bibinfo {author} {\bibfnamefont {A.~G.}\ \bibnamefont
  {Grushin}}, \bibinfo {author} {\bibfnamefont {T.}~\bibnamefont {Morimoto}},\
  and\ \bibinfo {author} {\bibfnamefont {J.~E.}\ \bibnamefont {Moore}},\ }\href
  {https://doi.org/10.1038/ncomms15995} {\bibfield  {journal} {\bibinfo
  {journal} {Nature Communications}\ }\textbf {\bibinfo {volume} {8}},\
  \bibinfo {pages} {15995} (\bibinfo {year} {2017})}\BibitemShut {NoStop}%
\bibitem [{\citenamefont {Rostami}\ and\ \citenamefont
  {Polini}(2018)}]{rostami2018}%
  \BibitemOpen
  \bibfield  {author} {\bibinfo {author} {\bibfnamefont {H.}~\bibnamefont
  {Rostami}}\ and\ \bibinfo {author} {\bibfnamefont {M.}~\bibnamefont
  {Polini}},\ }\href {https://doi.org/10.1103/PhysRevB.97.195151} {\bibfield
  {journal} {\bibinfo  {journal} {Physical Review B}\ }\textbf {\bibinfo
  {volume} {97}},\ \bibinfo {pages} {195151} (\bibinfo {year}
  {2018})}\BibitemShut {NoStop}%
\bibitem [{\citenamefont {Fregoso}(2019)}]{fregoso2019}%
  \BibitemOpen
  \bibfield  {author} {\bibinfo {author} {\bibfnamefont {B.~M.}\ \bibnamefont
  {Fregoso}},\ }\href {https://doi.org/10.1103/PhysRevB.100.064301} {\bibfield
  {journal} {\bibinfo  {journal} {Physical Review B}\ }\textbf {\bibinfo
  {volume} {100}},\ \bibinfo {pages} {064301} (\bibinfo {year}
  {2019})}\BibitemShut {NoStop}%
\bibitem [{\citenamefont {Sipe}\ and\ \citenamefont
  {Shkrebtii}(2000)}]{sipe2000}%
  \BibitemOpen
  \bibfield  {author} {\bibinfo {author} {\bibfnamefont {J.~E.}\ \bibnamefont
  {Sipe}}\ and\ \bibinfo {author} {\bibfnamefont {A.~I.}\ \bibnamefont
  {Shkrebtii}},\ }\href {https://doi.org/10.1103/PhysRevB.61.5337} {\bibfield
  {journal} {\bibinfo  {journal} {Physical Review B}\ }\textbf {\bibinfo
  {volume} {61}},\ \bibinfo {pages} {5337} (\bibinfo {year}
  {2000})}\BibitemShut {NoStop}%
\bibitem [{\citenamefont {Xu}\ \emph {et~al.}(2021)\citenamefont {Xu},
  \citenamefont {Zhou}, \citenamefont {Wang},\ and\ \citenamefont
  {Li}}]{xu2021}%
  \BibitemOpen
  \bibfield  {author} {\bibinfo {author} {\bibfnamefont {H.}~\bibnamefont
  {Xu}}, \bibinfo {author} {\bibfnamefont {J.}~\bibnamefont {Zhou}}, \bibinfo
  {author} {\bibfnamefont {H.}~\bibnamefont {Wang}},\ and\ \bibinfo {author}
  {\bibfnamefont {J.}~\bibnamefont {Li}},\ }\href
  {https://doi.org/10.1103/PhysRevB.103.205417} {\bibfield  {journal} {\bibinfo
   {journal} {Physical Review B}\ }\textbf {\bibinfo {volume} {103}},\ \bibinfo
  {pages} {205417} (\bibinfo {year} {2021})}\BibitemShut {NoStop}%
\bibitem [{\citenamefont {Adamantopoulos}\ \emph {et~al.}(2022)\citenamefont
  {Adamantopoulos}, \citenamefont {Merte}, \citenamefont {Go}, \citenamefont
  {Freimuth}, \citenamefont {Bl\"ugel},\ and\ \citenamefont
  {Mokrousov}}]{adamantopoulos2022}%
  \BibitemOpen
  \bibfield  {author} {\bibinfo {author} {\bibfnamefont {T.}~\bibnamefont
  {Adamantopoulos}}, \bibinfo {author} {\bibfnamefont {M.}~\bibnamefont
  {Merte}}, \bibinfo {author} {\bibfnamefont {D.}~\bibnamefont {Go}}, \bibinfo
  {author} {\bibfnamefont {F.}~\bibnamefont {Freimuth}}, \bibinfo {author}
  {\bibfnamefont {S.}~\bibnamefont {Bl\"ugel}},\ and\ \bibinfo {author}
  {\bibfnamefont {Y.}~\bibnamefont {Mokrousov}},\ }\href
  {https://doi.org/10.1103/PhysRevResearch.4.043046} {\bibfield  {journal}
  {\bibinfo  {journal} {Physical Review Research}\ }\textbf {\bibinfo {volume}
  {4}},\ \bibinfo {pages} {043046} (\bibinfo {year} {2022})}\BibitemShut
  {NoStop}%
\bibitem [{\citenamefont {Mancini}\ \emph {et~al.}(2015)\citenamefont
  {Mancini}, \citenamefont {Pagano}, \citenamefont {Cappellini}, \citenamefont
  {Livi}, \citenamefont {Rider}, \citenamefont {Catani}, \citenamefont {Sias},
  \citenamefont {Zoller}, \citenamefont {Inguscio}, \citenamefont {Dalmonte},\
  and\ \citenamefont {Fallani}}]{mancini2015}%
  \BibitemOpen
  \bibfield  {author} {\bibinfo {author} {\bibfnamefont {M.}~\bibnamefont
  {Mancini}}, \bibinfo {author} {\bibfnamefont {G.}~\bibnamefont {Pagano}},
  \bibinfo {author} {\bibfnamefont {G.}~\bibnamefont {Cappellini}}, \bibinfo
  {author} {\bibfnamefont {L.}~\bibnamefont {Livi}}, \bibinfo {author}
  {\bibfnamefont {M.}~\bibnamefont {Rider}}, \bibinfo {author} {\bibfnamefont
  {J.}~\bibnamefont {Catani}}, \bibinfo {author} {\bibfnamefont
  {C.}~\bibnamefont {Sias}}, \bibinfo {author} {\bibfnamefont {P.}~\bibnamefont
  {Zoller}}, \bibinfo {author} {\bibfnamefont {M.}~\bibnamefont {Inguscio}},
  \bibinfo {author} {\bibfnamefont {M.}~\bibnamefont {Dalmonte}},\ and\
  \bibinfo {author} {\bibfnamefont {L.}~\bibnamefont {Fallani}},\ }\href
  {https://doi.org/10.1126/science.aaa8736} {\bibfield  {journal} {\bibinfo
  {journal} {Science}\ }\textbf {\bibinfo {volume} {349}},\ \bibinfo {pages}
  {1510} (\bibinfo {year} {2015})}\BibitemShut {NoStop}%
\bibitem [{\citenamefont {Töpler}\ \emph {et~al.}(2021)\citenamefont
  {Töpler}, \citenamefont {Henk},\ and\ \citenamefont {Mertig}}]{Toepler2021}%
  \BibitemOpen
  \bibfield  {author} {\bibinfo {author} {\bibfnamefont {F.}~\bibnamefont
  {Töpler}}, \bibinfo {author} {\bibfnamefont {J.}~\bibnamefont {Henk}},\ and\
  \bibinfo {author} {\bibfnamefont {I.}~\bibnamefont {Mertig}},\ }\href
  {https://doi.org/10.1088/1367-2630/abe72b} {\bibfield  {journal} {\bibinfo
  {journal} {New Journal of Physics}\ }\textbf {\bibinfo {volume} {23}},\
  \bibinfo {pages} {033042} (\bibinfo {year} {2021})}\BibitemShut {NoStop}%
\bibitem [{\citenamefont {Ziolkowski}\ \emph {et~al.}(2023)\citenamefont
  {Ziolkowski}, \citenamefont {Busch}, \citenamefont {Mertig},\ and\
  \citenamefont {Henk}}]{Ziolkowski23}%
  \BibitemOpen
  \bibfield  {author} {\bibinfo {author} {\bibfnamefont {F.}~\bibnamefont
  {Ziolkowski}}, \bibinfo {author} {\bibfnamefont {O.}~\bibnamefont {Busch}},
  \bibinfo {author} {\bibfnamefont {I.}~\bibnamefont {Mertig}},\ and\ \bibinfo
  {author} {\bibfnamefont {J.}~\bibnamefont {Henk}},\ }\href
  {https://doi.org/10.1088/1361-648X/acb479} {\bibfield  {journal} {\bibinfo
  {journal} {Journal of Physics: Condensed Matter}\ }\textbf {\bibinfo {volume}
  {35}},\ \bibinfo {pages} {125501} (\bibinfo {year} {2023})}\BibitemShut
  {NoStop}%
\bibitem [{\citenamefont {Busch}\ \emph
  {et~al.}(2023{\natexlab{a}})\citenamefont {Busch}, \citenamefont
  {Ziolkowski}, \citenamefont {Mertig},\ and\ \citenamefont
  {Henk}}]{busch2023a}%
  \BibitemOpen
  \bibfield  {author} {\bibinfo {author} {\bibfnamefont {O.}~\bibnamefont
  {Busch}}, \bibinfo {author} {\bibfnamefont {F.}~\bibnamefont {Ziolkowski}},
  \bibinfo {author} {\bibfnamefont {I.}~\bibnamefont {Mertig}},\ and\ \bibinfo
  {author} {\bibfnamefont {J.}~\bibnamefont {Henk}},\ }\href
  {https://doi.org/https://doi.org/10.48550/arXiv.2303.09291} {\bibinfo {title}
  {Ultrafast dynamics of electrons excited by femtosecond laser pulses: spin
  polarization and spin-polarized currents}} (\bibinfo {year}
  {2023}{\natexlab{a}}),\ \Eprint {https://arxiv.org/abs/2303.09291}
  {arXiv:2303.09291 [cond-mat.mtrl-sci]} \BibitemShut {NoStop}%
\bibitem [{\citenamefont {Busch}\ \emph
  {et~al.}(2023{\natexlab{b}})\citenamefont {Busch}, \citenamefont
  {Ziolkowski}, \citenamefont {Mertig},\ and\ \citenamefont
  {Henk}}]{busch2023b}%
  \BibitemOpen
  \bibfield  {author} {\bibinfo {author} {\bibfnamefont {O.}~\bibnamefont
  {Busch}}, \bibinfo {author} {\bibfnamefont {F.}~\bibnamefont {Ziolkowski}},
  \bibinfo {author} {\bibfnamefont {I.}~\bibnamefont {Mertig}},\ and\ \bibinfo
  {author} {\bibfnamefont {J.}~\bibnamefont {Henk}},\ }\href
  {https://doi.org/https://doi.org/10.48550/arXiv.2306.12810} {\bibinfo {title}
  {Ultrafast dynamics of orbital angular momentum of electrons induced by
  femtosecond laser pulses: Generation and transfer across interfaces}}
  (\bibinfo {year} {2023}{\natexlab{b}}),\ \Eprint
  {https://arxiv.org/abs/2306.12810} {arXiv:2306.12810 [cond-mat.mtrl-sci]}
  \BibitemShut {NoStop}%
\bibitem [{Sup()}]{Supplement}%
  \BibitemOpen
  \href@noop {} {}\bibinfo {note} {See Supplemental Material at [URL will be
  inserted by publisher] for supporting information and additional
  results.}\BibitemShut {Stop}%
\bibitem [{\citenamefont {Slater}\ and\ \citenamefont
  {Koster}(1954)}]{Slater1954}%
  \BibitemOpen
  \bibfield  {author} {\bibinfo {author} {\bibfnamefont {J.~C.}\ \bibnamefont
  {Slater}}\ and\ \bibinfo {author} {\bibfnamefont {G.~F.}\ \bibnamefont
  {Koster}},\ }\href {https://doi.org/10.1103/PhysRev.94.1498} {\bibfield
  {journal} {\bibinfo  {journal} {Physical Review}\ }\textbf {\bibinfo {volume}
  {94}},\ \bibinfo {pages} {1498} (\bibinfo {year} {1954})}\BibitemShut
  {NoStop}%
\bibitem [{\citenamefont
  {Papaconstantopoulos}(2015)}]{Papaconstantopoulos2015}%
  \BibitemOpen
  \bibfield  {author} {\bibinfo {author} {\bibfnamefont {D.~A.}\ \bibnamefont
  {Papaconstantopoulos}},\ }\href
  {https://doi.org/https://doi.org/10.1007/978-1-4419-8264-3} {\emph {\bibinfo
  {title} {Handbook of the Band Structure of Elemental Solids}}}\ (\bibinfo
  {publisher} {Springer},\ \bibinfo {address} {Berlin},\ \bibinfo {year}
  {2015})\BibitemShut {NoStop}%
\bibitem [{\citenamefont {Konschuh}\ \emph {et~al.}(2010)\citenamefont
  {Konschuh}, \citenamefont {Gmitra},\ and\ \citenamefont
  {Fabian}}]{Konschuh2010}%
  \BibitemOpen
  \bibfield  {author} {\bibinfo {author} {\bibfnamefont {S.}~\bibnamefont
  {Konschuh}}, \bibinfo {author} {\bibfnamefont {M.}~\bibnamefont {Gmitra}},\
  and\ \bibinfo {author} {\bibfnamefont {J.}~\bibnamefont {Fabian}},\ }\href
  {https://doi.org/10.1103/PhysRevB.82.245412} {\bibfield  {journal} {\bibinfo
  {journal} {Physical Review B}\ }\textbf {\bibinfo {volume} {82}},\ \bibinfo
  {pages} {245412} (\bibinfo {year} {2010})}\BibitemShut {NoStop}%
\bibitem [{\citenamefont {Savasta}\ and\ \citenamefont
  {Girlanda}(1995)}]{Savasta1995}%
  \BibitemOpen
  \bibfield  {author} {\bibinfo {author} {\bibfnamefont {S.}~\bibnamefont
  {Savasta}}\ and\ \bibinfo {author} {\bibfnamefont {R.}~\bibnamefont
  {Girlanda}},\ }\href
  {https://doi.org/https://doi.org/10.1016/0038-1098(95)00242-1} {\bibfield
  {journal} {\bibinfo  {journal} {Solid State Communications}\ }\textbf
  {\bibinfo {volume} {96}},\ \bibinfo {pages} {517} (\bibinfo {year}
  {1995})}\BibitemShut {NoStop}%
\bibitem [{\citenamefont {Henk}\ \emph {et~al.}(1996)\citenamefont {Henk},
  \citenamefont {Scheunemann}, \citenamefont {Halilov},\ and\ \citenamefont
  {Feder}}]{Henk1996}%
  \BibitemOpen
  \bibfield  {author} {\bibinfo {author} {\bibfnamefont {J.}~\bibnamefont
  {Henk}}, \bibinfo {author} {\bibfnamefont {T.}~\bibnamefont {Scheunemann}},
  \bibinfo {author} {\bibfnamefont {S.~V.}\ \bibnamefont {Halilov}},\ and\
  \bibinfo {author} {\bibfnamefont {R.}~\bibnamefont {Feder}},\ }\href
  {https://doi.org/10.1088/0953-8984/8/1/007} {\bibfield  {journal} {\bibinfo
  {journal} {Journal of Physics: Condensed Matter}\ }\textbf {\bibinfo {volume}
  {8}},\ \bibinfo {pages} {47} (\bibinfo {year} {1996})}\BibitemShut {NoStop}%
\bibitem [{\citenamefont {Mahan}(2000)}]{mahan2013}%
  \BibitemOpen
  \bibfield  {author} {\bibinfo {author} {\bibfnamefont {G.~D.}\ \bibnamefont
  {Mahan}},\ }\href {https://doi.org/https://doi.org/10.1007/978-1-4757-5714-9}
  {\emph {\bibinfo {title} {Many-Particle Physics}}},\ \bibinfo {edition}
  {3rd}\ ed.\ (\bibinfo  {publisher} {Springer},\ \bibinfo {address} {New
  York},\ \bibinfo {year} {2000})\BibitemShut {NoStop}%
\bibitem [{\citenamefont {Pezo}\ \emph {et~al.}(2022)\citenamefont {Pezo},
  \citenamefont {Garc\'{\i}a~Ovalle},\ and\ \citenamefont
  {Manchon}}]{pezo2022}%
  \BibitemOpen
  \bibfield  {author} {\bibinfo {author} {\bibfnamefont {A.}~\bibnamefont
  {Pezo}}, \bibinfo {author} {\bibfnamefont {D.}~\bibnamefont
  {Garc\'{\i}a~Ovalle}},\ and\ \bibinfo {author} {\bibfnamefont
  {A.}~\bibnamefont {Manchon}},\ }\href
  {https://doi.org/10.1103/PhysRevB.106.104414} {\bibfield  {journal} {\bibinfo
   {journal} {Physical Review B}\ }\textbf {\bibinfo {volume} {106}},\ \bibinfo
  {pages} {104414} (\bibinfo {year} {2022})}\BibitemShut {NoStop}%
\bibitem [{Note1()}]{Note1}%
  \BibitemOpen
  \bibinfo {note} {Focusing on nonequilibrium scenarios, possible equilibrium
  currents are subtracted. The latter would vanish if angular-momentum currents
  were properly defined; see Refs.~\protect \rev@citealp {sun2005, shi2006} for
  a discussion of spin currents. Explicit expressions for the currents are
  given in the Supplemental Material~\cite {Supplement}.}\BibitemShut {Stop}%
\bibitem [{\citenamefont {Niimi}\ \emph {et~al.}(2011)\citenamefont {Niimi},
  \citenamefont {Morota}, \citenamefont {Wei}, \citenamefont {Deranlot},
  \citenamefont {Basletic}, \citenamefont {Hamzic}, \citenamefont {Fert},\ and\
  \citenamefont {Otani}}]{niimi2011}%
  \BibitemOpen
  \bibfield  {author} {\bibinfo {author} {\bibfnamefont {Y.}~\bibnamefont
  {Niimi}}, \bibinfo {author} {\bibfnamefont {M.}~\bibnamefont {Morota}},
  \bibinfo {author} {\bibfnamefont {D.~H.}\ \bibnamefont {Wei}}, \bibinfo
  {author} {\bibfnamefont {C.}~\bibnamefont {Deranlot}}, \bibinfo {author}
  {\bibfnamefont {M.}~\bibnamefont {Basletic}}, \bibinfo {author}
  {\bibfnamefont {A.}~\bibnamefont {Hamzic}}, \bibinfo {author} {\bibfnamefont
  {A.}~\bibnamefont {Fert}},\ and\ \bibinfo {author} {\bibfnamefont
  {Y.}~\bibnamefont {Otani}},\ }\href
  {https://doi.org/10.1103/PhysRevLett.106.126601} {\bibfield  {journal}
  {\bibinfo  {journal} {Physical Review Letters}\ }\textbf {\bibinfo {volume}
  {106}},\ \bibinfo {pages} {126601} (\bibinfo {year} {2011})}\BibitemShut
  {NoStop}%
\bibitem [{\citenamefont {Ding}\ \emph {et~al.}(2020)\citenamefont {Ding},
  \citenamefont {Ross}, \citenamefont {Go}, \citenamefont {Baldrati},
  \citenamefont {Ren}, \citenamefont {Freimuth}, \citenamefont {Becker},
  \citenamefont {Kammerbauer}, \citenamefont {Yang}, \citenamefont {Jakob},
  \citenamefont {Mokrousov},\ and\ \citenamefont {Kl\"aui}}]{ding2020}%
  \BibitemOpen
  \bibfield  {author} {\bibinfo {author} {\bibfnamefont {S.}~\bibnamefont
  {Ding}}, \bibinfo {author} {\bibfnamefont {A.}~\bibnamefont {Ross}}, \bibinfo
  {author} {\bibfnamefont {D.}~\bibnamefont {Go}}, \bibinfo {author}
  {\bibfnamefont {L.}~\bibnamefont {Baldrati}}, \bibinfo {author}
  {\bibfnamefont {Z.}~\bibnamefont {Ren}}, \bibinfo {author} {\bibfnamefont
  {F.}~\bibnamefont {Freimuth}}, \bibinfo {author} {\bibfnamefont
  {S.}~\bibnamefont {Becker}}, \bibinfo {author} {\bibfnamefont
  {F.}~\bibnamefont {Kammerbauer}}, \bibinfo {author} {\bibfnamefont
  {J.}~\bibnamefont {Yang}}, \bibinfo {author} {\bibfnamefont {G.}~\bibnamefont
  {Jakob}}, \bibinfo {author} {\bibfnamefont {Y.}~\bibnamefont {Mokrousov}},\
  and\ \bibinfo {author} {\bibfnamefont {M.}~\bibnamefont {Kl\"aui}},\ }\href
  {https://doi.org/10.1103/PhysRevLett.125.177201} {\bibfield  {journal}
  {\bibinfo  {journal} {Physical Review Letters}\ }\textbf {\bibinfo {volume}
  {125}},\ \bibinfo {pages} {177201} (\bibinfo {year} {2020})}\BibitemShut
  {NoStop}%
\bibitem [{\citenamefont {Go}\ and\ \citenamefont {Lee}(2020)}]{go2020}%
  \BibitemOpen
  \bibfield  {author} {\bibinfo {author} {\bibfnamefont {D.}~\bibnamefont
  {Go}}\ and\ \bibinfo {author} {\bibfnamefont {H.-W.}\ \bibnamefont {Lee}},\
  }\href {https://doi.org/10.1103/PhysRevResearch.2.013177} {\bibfield
  {journal} {\bibinfo  {journal} {Physical Review Research}\ }\textbf {\bibinfo
  {volume} {2}},\ \bibinfo {pages} {013177} (\bibinfo {year}
  {2020})}\BibitemShut {NoStop}%
\bibitem [{\citenamefont {Sala}\ and\ \citenamefont
  {Gambardella}(2022)}]{sala2022}%
  \BibitemOpen
  \bibfield  {author} {\bibinfo {author} {\bibfnamefont {G.}~\bibnamefont
  {Sala}}\ and\ \bibinfo {author} {\bibfnamefont {P.}~\bibnamefont
  {Gambardella}},\ }\href {https://doi.org/10.1103/PhysRevResearch.4.033037}
  {\bibfield  {journal} {\bibinfo  {journal} {Physical Review Research}\
  }\textbf {\bibinfo {volume} {4}},\ \bibinfo {pages} {033037} (\bibinfo {year}
  {2022})}\BibitemShut {NoStop}%
\bibitem [{\citenamefont {Busch}\ \emph
  {et~al.}(2023{\natexlab{c}})\citenamefont {Busch}, \citenamefont {Mertig},\
  and\ \citenamefont {Göbel}}]{busch2023c}%
  \BibitemOpen
  \bibfield  {author} {\bibinfo {author} {\bibfnamefont {O.}~\bibnamefont
  {Busch}}, \bibinfo {author} {\bibfnamefont {I.}~\bibnamefont {Mertig}},\ and\
  \bibinfo {author} {\bibfnamefont {B.}~\bibnamefont {Göbel}},\ }\href
  {https://doi.org/https://doi.org/10.48550/arXiv.2306.17295} {\bibinfo {title}
  {Orbital hall effect and orbital edge states caused by {$s$} electrons}}
  (\bibinfo {year} {2023}{\natexlab{c}}),\ \Eprint
  {https://arxiv.org/abs/2306.17295} {arXiv:2306.17295 [cond-mat.mes-hall]}
  \BibitemShut {NoStop}%
\bibitem [{\citenamefont {Canonico}\ \emph
  {et~al.}(2020{\natexlab{a}})\citenamefont {Canonico}, \citenamefont {Cysne},
  \citenamefont {Molina-Sanchez}, \citenamefont {Muniz},\ and\ \citenamefont
  {Rappoport}}]{canonico2020a}%
  \BibitemOpen
  \bibfield  {author} {\bibinfo {author} {\bibfnamefont {L.~M.}\ \bibnamefont
  {Canonico}}, \bibinfo {author} {\bibfnamefont {T.~P.}\ \bibnamefont {Cysne}},
  \bibinfo {author} {\bibfnamefont {A.}~\bibnamefont {Molina-Sanchez}},
  \bibinfo {author} {\bibfnamefont {R.~B.}\ \bibnamefont {Muniz}},\ and\
  \bibinfo {author} {\bibfnamefont {T.~G.}\ \bibnamefont {Rappoport}},\ }\href
  {https://doi.org/10.1103/PhysRevB.101.161409} {\bibfield  {journal} {\bibinfo
   {journal} {Physical Review B}\ }\textbf {\bibinfo {volume} {101}},\ \bibinfo
  {pages} {161409} (\bibinfo {year} {2020}{\natexlab{a}})}\BibitemShut
  {NoStop}%
\bibitem [{\citenamefont {Canonico}\ \emph
  {et~al.}(2020{\natexlab{b}})\citenamefont {Canonico}, \citenamefont {Cysne},
  \citenamefont {Rappoport},\ and\ \citenamefont {Muniz}}]{canonico2020b}%
  \BibitemOpen
  \bibfield  {author} {\bibinfo {author} {\bibfnamefont {L.~M.}\ \bibnamefont
  {Canonico}}, \bibinfo {author} {\bibfnamefont {T.~P.}\ \bibnamefont {Cysne}},
  \bibinfo {author} {\bibfnamefont {T.~G.}\ \bibnamefont {Rappoport}},\ and\
  \bibinfo {author} {\bibfnamefont {R.~B.}\ \bibnamefont {Muniz}},\ }\href
  {https://doi.org/10.1103/PhysRevB.101.075429} {\bibfield  {journal} {\bibinfo
   {journal} {Physical Review B}\ }\textbf {\bibinfo {volume} {101}},\ \bibinfo
  {pages} {075429} (\bibinfo {year} {2020}{\natexlab{b}})}\BibitemShut
  {NoStop}%
\bibitem [{\citenamefont {Salemi}\ and\ \citenamefont
  {Oppeneer}(2022)}]{salemi2023}%
  \BibitemOpen
  \bibfield  {author} {\bibinfo {author} {\bibfnamefont {L.}~\bibnamefont
  {Salemi}}\ and\ \bibinfo {author} {\bibfnamefont {P.~M.}\ \bibnamefont
  {Oppeneer}},\ }\href {https://doi.org/10.1103/PhysRevMaterials.6.095001}
  {\bibfield  {journal} {\bibinfo  {journal} {Physical Review Materials}\
  }\textbf {\bibinfo {volume} {6}},\ \bibinfo {pages} {095001} (\bibinfo {year}
  {2022})}\BibitemShut {NoStop}%
\bibitem [{\citenamefont {Seifert}\ \emph {et~al.}(2023)\citenamefont
  {Seifert}, \citenamefont {Go}, \citenamefont {Hayashi}, \citenamefont
  {Rouzegar}, \citenamefont {Freimuth}, \citenamefont {Ando}, \citenamefont
  {Mokrousov},\ and\ \citenamefont {Kampfrath}}]{seifert2023}%
  \BibitemOpen
  \bibfield  {author} {\bibinfo {author} {\bibfnamefont {T.~S.}\ \bibnamefont
  {Seifert}}, \bibinfo {author} {\bibfnamefont {D.}~\bibnamefont {Go}},
  \bibinfo {author} {\bibfnamefont {H.}~\bibnamefont {Hayashi}}, \bibinfo
  {author} {\bibfnamefont {R.}~\bibnamefont {Rouzegar}}, \bibinfo {author}
  {\bibfnamefont {F.}~\bibnamefont {Freimuth}}, \bibinfo {author}
  {\bibfnamefont {K.}~\bibnamefont {Ando}}, \bibinfo {author} {\bibfnamefont
  {Y.}~\bibnamefont {Mokrousov}},\ and\ \bibinfo {author} {\bibfnamefont
  {T.}~\bibnamefont {Kampfrath}},\ }\href
  {https://doi.org/https://doi.org/10.48550/arXiv.2301.00747} {\bibinfo {title}
  {Time-domain observation of ballistic orbital-angular-momentum currents with
  giant relaxation length in tungsten}} (\bibinfo {year} {2023}),\ \Eprint
  {https://arxiv.org/abs/2301.00747} {arXiv:2301.00747 [physics.app-ph]}
  \BibitemShut {NoStop}%
\bibitem [{\citenamefont {Sun}\ and\ \citenamefont {Xie}(2005)}]{sun2005}%
  \BibitemOpen
  \bibfield  {author} {\bibinfo {author} {\bibfnamefont {Q.-f.}\ \bibnamefont
  {Sun}}\ and\ \bibinfo {author} {\bibfnamefont {X.~C.}\ \bibnamefont {Xie}},\
  }\href {https://doi.org/10.1103/PhysRevB.72.245305} {\bibfield  {journal}
  {\bibinfo  {journal} {Physical Review B}\ }\textbf {\bibinfo {volume} {72}},\
  \bibinfo {pages} {245305} (\bibinfo {year} {2005})}\BibitemShut {NoStop}%
\bibitem [{\citenamefont {Shi}\ \emph {et~al.}(2006)\citenamefont {Shi},
  \citenamefont {Zhang}, \citenamefont {Xiao},\ and\ \citenamefont
  {Niu}}]{shi2006}%
  \BibitemOpen
  \bibfield  {author} {\bibinfo {author} {\bibfnamefont {J.}~\bibnamefont
  {Shi}}, \bibinfo {author} {\bibfnamefont {P.}~\bibnamefont {Zhang}}, \bibinfo
  {author} {\bibfnamefont {D.}~\bibnamefont {Xiao}},\ and\ \bibinfo {author}
  {\bibfnamefont {Q.}~\bibnamefont {Niu}},\ }\href
  {https://doi.org/10.1103/PhysRevLett.96.076604} {\bibfield  {journal}
  {\bibinfo  {journal} {Physical Review Letters}\ }\textbf {\bibinfo {volume}
  {96}},\ \bibinfo {pages} {076604} (\bibinfo {year} {2006})}\BibitemShut
  {NoStop}%
\bibitem [{\citenamefont {Nikoli{\'c}}\ \emph {et~al.}(2006)\citenamefont
  {Nikoli{\'c}}, \citenamefont {Z{\^a}rbo},\ and\ \citenamefont
  {Souma}}]{Nikolic2006}%
  \BibitemOpen
  \bibfield  {author} {\bibinfo {author} {\bibfnamefont {B.~K.}\ \bibnamefont
  {Nikoli{\'c}}}, \bibinfo {author} {\bibfnamefont {L.~P.}\ \bibnamefont
  {Z{\^a}rbo}},\ and\ \bibinfo {author} {\bibfnamefont {S.}~\bibnamefont
  {Souma}},\ }\href {https://doi.org/10.1103/PhysRevB.73.075303} {\bibfield
  {journal} {\bibinfo  {journal} {Physical Review B}\ }\textbf {\bibinfo
  {volume} {73}},\ \bibinfo {pages} {075303} (\bibinfo {year}
  {2006})}\BibitemShut {NoStop}%
\bibitem [{\citenamefont {Petrovi{\'c}}\ \emph {et~al.}(2018)\citenamefont
  {Petrovi{\'c}}, \citenamefont {Popescu}, \citenamefont {Bajpai},
  \citenamefont {Plech{\'a}{\v{c}}},\ and\ \citenamefont
  {Nikoli{\'c}}}]{Petrovic2018}%
  \BibitemOpen
  \bibfield  {author} {\bibinfo {author} {\bibfnamefont {M.~D.}\ \bibnamefont
  {Petrovi{\'c}}}, \bibinfo {author} {\bibfnamefont {B.~S.}\ \bibnamefont
  {Popescu}}, \bibinfo {author} {\bibfnamefont {U.}~\bibnamefont {Bajpai}},
  \bibinfo {author} {\bibfnamefont {P.}~\bibnamefont {Plech{\'a}{\v{c}}}},\
  and\ \bibinfo {author} {\bibfnamefont {B.~K.}\ \bibnamefont {Nikoli{\'c}}},\
  }\href {https://doi.org/10.1103/PhysRevApplied.10.054038} {\bibfield
  {journal} {\bibinfo  {journal} {Physical Review Applied}\ }\textbf {\bibinfo
  {volume} {10}},\ \bibinfo {pages} {054038} (\bibinfo {year}
  {2018})}\BibitemShut {NoStop}%
\bibitem [{\citenamefont {Jo}\ \emph {et~al.}(2018)\citenamefont {Jo},
  \citenamefont {Go},\ and\ \citenamefont {Lee}}]{daegeun2018}%
  \BibitemOpen
  \bibfield  {author} {\bibinfo {author} {\bibfnamefont {D.}~\bibnamefont
  {Jo}}, \bibinfo {author} {\bibfnamefont {D.}~\bibnamefont {Go}},\ and\
  \bibinfo {author} {\bibfnamefont {H.-W.}\ \bibnamefont {Lee}},\ }\href
  {https://doi.org/10.1103/PhysRevB.98.214405} {\bibfield  {journal} {\bibinfo
  {journal} {Physical Review B}\ }\textbf {\bibinfo {volume} {98}},\ \bibinfo
  {pages} {214405} (\bibinfo {year} {2018})}\BibitemShut {NoStop}%
\bibitem [{\citenamefont {Rashba}(2003)}]{rashba2003}%
  \BibitemOpen
  \bibfield  {author} {\bibinfo {author} {\bibfnamefont {E.~I.}\ \bibnamefont
  {Rashba}},\ }\href {https://doi.org/10.1103/PhysRevB.68.241315} {\bibfield
  {journal} {\bibinfo  {journal} {Physical Review B}\ }\textbf {\bibinfo
  {volume} {68}},\ \bibinfo {pages} {241315} (\bibinfo {year}
  {2003})}\BibitemShut {NoStop}%
\bibitem [{\citenamefont {Adagideli}\ \emph {et~al.}(2007)\citenamefont
  {Adagideli}, \citenamefont {Scheid}, \citenamefont {Wimmer}, \citenamefont
  {Bauer},\ and\ \citenamefont {Richter}}]{adagideli2007}%
  \BibitemOpen
  \bibfield  {author} {\bibinfo {author} {\bibfnamefont {{\.I}.}~\bibnamefont
  {Adagideli}}, \bibinfo {author} {\bibfnamefont {M.}~\bibnamefont {Scheid}},
  \bibinfo {author} {\bibfnamefont {M.}~\bibnamefont {Wimmer}}, \bibinfo
  {author} {\bibfnamefont {G.~E.~W.}\ \bibnamefont {Bauer}},\ and\ \bibinfo
  {author} {\bibfnamefont {K.}~\bibnamefont {Richter}},\ }\href
  {https://doi.org/10.1088/1367-2630/9/10/382} {\bibfield  {journal} {\bibinfo
  {journal} {New Journal of Physics}\ }\textbf {\bibinfo {volume} {9}},\
  \bibinfo {pages} {382} (\bibinfo {year} {2007})}\BibitemShut {NoStop}%
\bibitem [{\citenamefont {Tokatly}(2008)}]{tokatly2008}%
  \BibitemOpen
  \bibfield  {author} {\bibinfo {author} {\bibfnamefont {I.~V.}\ \bibnamefont
  {Tokatly}},\ }\href {https://doi.org/10.1103/PhysRevLett.101.106601}
  {\bibfield  {journal} {\bibinfo  {journal} {Physical Review Letters}\
  }\textbf {\bibinfo {volume} {101}},\ \bibinfo {pages} {106601} (\bibinfo
  {year} {2008})}\BibitemShut {NoStop}%
\end{thebibliography}%

% For arXiv, input the supplement here => Comment/Uncomment this part
\clearpage
\begin{widetext}
\section*{Supplemental Material}
\section{I. Ultrafast electron dynamics}
\label{sec:electron_dynamics}
The samples are free-standing fcc(001) monolayers of $15$~atomic rows width (`across the ribbon'). These layers form a square lattice, with Cartesian axis chosen as $x \equiv [110]$, $y \equiv [\bar{1}10]$, and $z \equiv [001]$. We apply periodic boundary conditions in $x$ direction (`along the ribbon').

The electronic structure of the samples is described by a tight-binding Hamiltonian~$\operator{H}_0$ of Slater-Koster type~\cite{Slater1954}, with parameters for the s-, p-, and d-orbitals taken from Ref.~\onlinecite{Papaconstantopoulos2015}. Spin-orbit coupling is taken into account as described in Ref.~\onlinecite{Konschuh2010}.

The electron system is excited by a femtosecond laser pulse with photon energy $E_{\mathrm{ph}} = \hbar\omega$. In the present work, the electromagnetic radiation impinges along the $z$-axis onto the ribbon (cf.~Figure~1 in the main manuscript), and the azimuth of incidence is chosen such that the electric field oscillates along the $x$ axis. Hence, the laser's electric field is written as
\begin{align}
    \vec{E}(t) & = E_{0} \, l(t) \cos(\omega t) \, \vec{e}_{x}.
    \label{eq:e-field}
\end{align}
Here,  $l(t)$ is a Lorentzian envelope, $E_{0}$ the amplitude, and $\vec{e}_{x}$ the unit vector in $x$-direction. The pulse shape $l(t) \cos(\omega t)$ sets the timeframe of the simulations and is thus shown schematically in Fig.~2 of the main text.

The electron dynamics is described by the von Neumann equation
\begin{align}
   -\cone \hbar \frac{\mathrm{d} \operator{\rho}(t)}{\mathrm{d} t} & =  [ \operator{\rho}(t),\operator{H}(t)]
    \label{eq:EOMSupp}
\end{align}
for the one-particle density matrix
\begin{align}
    \operator{\rho}(t) = \sum_{n, m} \ket{n} \, p_{nm}(t) \, \bra{m}.
\end{align}
$\{ \ket{n} \}$ is the set of eigenstates of $\operator{H}_0$, with $\operator{H}_0 \ket{n} = \epsilon_{n} \ket{n}$. The time-dependent Hamiltonian $\operator{H}(t)$ comprises the electric field of the laser via minimal coupling~\cite{Savasta1995}. The equation of motion~\eqref{eq:EOMSupp} for $\operator{\rho}(t)$ is solved within our theoretical framework \textsc{evolve}; for details see Ref.~\onlinecite{Toepler2021}. 

\section{II. Orbital angular momentum currents}
Similarly to the currents of spin angular momentum (SAM;  Ref.~\onlinecite{busch2023a}), currents of orbital angular momentum (OAM) are derived from the symmetrized expression
\begin{align}
	\expect{j_{kl}^{\mu}}(t) & \equiv \frac{1}{2} \left[ \expect{L^{\mu} j_{kl}}(t) + \expect{j_{kl} L^{\mu}}(t) \right], \quad \mu = x, y, z
    \label{eq:Mahan}
\end{align}
(see for example Refs.~\onlinecite{Nikolic2006, Petrovic2018}), in which the operator $\operator{\jmath}_{kl}$ for the current from site $l$ to site $k$ is adapted from Mahan's expression~\cite{mahan2013}. 
Using a tight-binding approach for the electronic structure, the one-particle density matrix and the Hamiltonian are given in matrix form. Both are represented in a site-orbital-spin basis $\{ \ket{ k \alpha \sigma} \}$ ($k$ site index, $\alpha$ orbital, $\sigma$ spin orientation with respect to the $z$ axis). This allows to define block matrices
\begin{align}
\left( \mat{p}_{kl}^{\sigma \sigma'} \right)_{\alpha \beta} = p_{k \alpha \sigma, l \beta \sigma'},
\quad
\left( \mat{t}_{kl}^{\sigma \sigma'} \right)_{\alpha \beta} = t_{k \alpha \sigma, l \beta \sigma'}
\end{align}
that are indexed with respect to site and spin. These are further combined into site-indexed $2 \times 2$ block matrices,
\begin{align}
\mat{P}_{lk} & = 
    \begin{pmatrix}
    \mat{p}_{lk}^{\uparrow \uparrow} & \mat{p}_{lk}^{\uparrow \downarrow} \\
    \mat{p}_{lk}^{\downarrow \uparrow} & \mat{p}_{lk}^{\downarrow \downarrow}
    \end{pmatrix},
\quad
\mat{T}_{kl} = 
    \begin{pmatrix}
    \mat{t}_{kl}^{\uparrow \uparrow} & \mat{t}_{kl}^{\uparrow \downarrow} \\
    \mat{t}_{kl}^{\downarrow \uparrow} & \mat{t}_{kl}^{\downarrow \downarrow}
    \end{pmatrix}.
\end{align}
With this, \eqref{eq:Mahan} can be written as
\begin{align}
\expect{j_{kl}^{\mu}} = & - \frac{\cone}{4} \operatorname{tr}
\mat{P}_{lk} \left[ \mat{L}^{\mu}, \mat{T}_{kl} \right]_{+}
- \expect{l \leftrightarrow k}, \quad \mu = x, y, z,
\label{eq:current-matrix-1}
\end{align}
in which $[ \cdot , \cdot]_{+}$ is the anticommutator.  The $\mu$-th component of the OAM operator appears as matrices
\begin{align}
\mat{L}^{\mu}_{m} = 
    \begin{pmatrix}
    \mat{l}^{\mu}_{m} & \mat{0} \\
    \mat{0} & \mat{l}^{\mu}_{m}
    \end{pmatrix},
\end{align}
which are diagonal with respect to both site and spin, but depend on the atomic species at site $m$. Hence, $\mat{L}^{\mu}$ in~\eqref{eq:current-matrix-1} is taken either at $m = k$ or at $m = l$. Explicit forms of $\mat{L}^{\mu}_{m}$ for cubic harmonics can be found for example in Ref.~\onlinecite{daegeun2018}.

With the above definitions, \eqref{eq:current-matrix-1} becomes
\begin{align}
    \expect{j_{kl}^{\mu}} = & - \frac{\cone}{4} \operatorname{tr} \left[
    \begin{pmatrix}
        \mat{p}_{lk}^{\uparrow \uparrow} & \mat{p}_{lk}^{\uparrow \downarrow} \\
        \mat{p}_{lk}^{\downarrow \uparrow} & \mat{p}_{lk}^{\downarrow \downarrow}
    \end{pmatrix}
    \left[
    \mat{l}^{\mu}_{k}
    \begin{pmatrix}
    \mat{t}_{kl}^{\uparrow \uparrow} & \mat{t}_{kl}^{\uparrow \downarrow} \\
    \mat{t}_{kl}^{\downarrow \uparrow} & \mat{t}_{kl}^{\downarrow \downarrow}
    \end{pmatrix}
    + 
    \begin{pmatrix}
    \mat{t}_{kl}^{\uparrow \uparrow} & \mat{t}_{kl}^{\uparrow \downarrow} \\
    \mat{t}_{kl}^{\downarrow \uparrow} & \mat{t}_{kl}^{\downarrow \downarrow}
    \end{pmatrix}
    \mat{l}^{\mu}_{\edge{l}}
    \right] \right] - \expect{l \leftrightarrow k}.
\end{align}
Performing the matrix multiplication leaves us with $\expect{j_{kl}^{\mu}} = \expect{j_{kl}^{\mu}}{}^{\sigma\sigma} + \expect{j_{kl}^{\mu}}{}^{\sigma\sigma'}$, that is a sum of `spin-conserving' ($\uparrow \uparrow$ or $\downarrow \downarrow$) and `spin-mixing' ($\uparrow \downarrow$ or $\downarrow \uparrow$) contributions.

Exploiting the hermiticity of the matrices and the invariance of the trace upon cyclic permutations,  the spin-conserving terms are condensed into
\begin{align}
    \expect{j_{kl}^{\mu}}^{\sigma\sigma} = \frac{1}{2} \mathrm{Im} \operatorname{tr} \left[
    \mat{p}_{lk}^{\uparrow \uparrow} 
    \left(
        \mat{l}^{\mu}_{k} \mat{t}_{kl}^{\uparrow \uparrow} 
        +
        \mat{t}_{kl}^{\uparrow \uparrow} \mat{l}^{\mu}_{\edge{l}}
    \right)
    + 
    \mat{p}_{lk}^{\downarrow \downarrow}
    \left(
        \mat{l}^{\mu}_{k} \mat{t}_{kl}^{\downarrow \downarrow} 
        +
        \mat{t}_{kl}^{\downarrow \downarrow} \mat{l}^{\mu}_{\edge{l}}
    \right)
    \right]
\end{align}
and the spin-mixing terms into
\begin{align}
    \expect{j_{kl}^{\mu}}^{\sigma\sigma'} = \frac{1}{2} \mathrm{Im}  \operatorname{tr} \left[ 
    \mat{p}_{lk}^{\uparrow \downarrow}
    \left(
        \mat{l}^{\mu}_{k} \mat{t}_{kl}^{\downarrow \uparrow} 
        +
        \mat{t}_{kl}^{\downarrow \uparrow} \mat{l}^{\mu}_{\edge{l}}
    \right)
    + 
    \mat{p}_{lk}^{\downarrow \uparrow} 
    \left(
        \mat{l}^{\mu}_{k} \mat{t}_{kl}^{\uparrow \downarrow} 
        +
        \mat{t}_{kl}^{\uparrow \downarrow} \mat{l}^{\mu}_{\edge{l}}
    \right)
    \right]
\end{align}
($\sigma = - \sigma'$). Adding both contributions, we end up with
\begin{align}
    \expect{j_{kl}^{\mu}} & = 
    \frac{1}{2} \mathrm{Im} \sum_{\sigma, \sigma'} 
    \operatorname{tr}
        \mat{p}_{lk}^{\sigma \sigma'}
        \left(
        \mat{l}^{\mu}_{k} \mat{t}_{kl}^{\sigma' \sigma} 
        +
        \mat{t}_{kl}^{\sigma' \sigma} \mat{l}^{\mu}_{\edge{l}}
        \right).
\end{align}
The probability current can be obtained from this expression by setting $\mat{l}^{\mu}_{k} = \mat{1}$,
\begin{align}
    \expect{j_{kl}} & = 
    \mathrm{Im} \sum_{\sigma, \sigma'} 
    \operatorname{tr}
      \mat{p}_{lk}^{\sigma \sigma'}
      \mat{t}_{kl}^{\sigma' \sigma}.
\end{align}

The currents obey the symmetries $\expect{j_{kl}} = - \expect{j_{lk}}$ and  $\expect{j_{kl}^{\mu}} = - \expect{j_{lk}^{\mu}} = - \expect{j_{kl}^{-\mu}}$. Moreover, they fulfill Kirchhoff's rules, as has been checked in numerical simulations.

We focus on nonequilibrium currents, i.e. more precisely on the laser-induced changes with respect to equilibrium currents~\cite{rashba2003, Nikolic2006, adagideli2007, tokatly2008} that may exist before the laser excitation.

\section{III. Symmetry analysis}
\label{sec:symmetry}

Instead of a full group-theoretical analysis~\cite{Henk1996}, we perform a symmetry analysis which tells what components of the laser-induced OAM are forbidden for the given setup. 

The electric field $E$ of the laser is oriented along the $x$-direction. Thus, for the chosen nanoribbon, symmetries that leave $E$ and the ribbon invariant are the reflection $\operator{m}_{y}$ at the $xz$ plane: $(x, y, z) \to (x, -y, z)$ and the $\pi$-rotation $\operator{C}_{2x}$: $(x, y, z) \to (x, -y, -z)$ about the central line along the nanoribbon (denoted as $x$-axis here).

Since we are interested in accumulation of OAM at the ribbon's edges, the OAM is decomposed into two regions of the nanoribbon, one with $y < 0$ (`lower', $\edge{l}$) and the other with $y > 0$ (`upper', $\edge{u}$). Both symmetry operations interchange the edges, that is $\edge{l} \leftrightarrow \edge{u}$.

\begin{table}[h!]
	\caption{Effect of symmetry operations (left column) components $L_{\edge{r}}^{\mu}$ of the orbital angular momentum in the `lower' region ($\edge{r} = \edge{l}$) and in the `upper' ($\edge{r} = \edge{u}$) of the nanoribbon ($\mu = x, y, z$). $\operator{1}$ is the identity operation, $\operator{m}_{y}$ and $\operator{C}_{2x}$ are defined in the text.}
	\centering
	\begin{tabular}{c|rrr|rrr}
		\hline \hline
		$\operator{1}$     &  $L_{\edge{l}}^{x}$  &   $L_{\edge{l}}^{y}$  &  $L_{\edge{l}}^{z}$ & $L_{\edge{u}}^{x}$  &   $L_{\edge{u}}^{y}$  &  $L_{\edge{u}}^{z}$ \\
		$\operator{m}_{y}$  &  $-L_{\edge{u}}^{x}$ &   $L_{\edge{u}}^{y}$  &  $-L_{\edge{u}}^{z}$ & $-L_{\edge{l}}^{x}$  &   $L_{\edge{l}}^{y}$  &  $-L_{\edge{l}}^{z}$\\
		$\operator{C}_{2x}$ &  $L_{\edge{u}}^{x}$ &   $-L_{\edge{u}}^{y}$  &  $-L_{\edge{u}}^{z}$ & $L_{\edge{l}}^{x}$  &   $-L_{\edge{l}}^{y}$  &  $-L_{\edge{l}}^{z}$ \\
		\hline \hline 
	\end{tabular}
    \label{tab:symmetry}
\end{table}

Inspection of Table~\ref{tab:symmetry} tells immediately that $L_{\edge{l}}^{x} = -L_{\edge{u}}^{x} = L_{\edge{u}}^{x}$ (second column) and $L_{\edge{u}}^{x} = -L_{\edge{l}}^{x} = L_{\edge{l}}^{x}$ (fifth column), which can only be fulfilled by $L_{\edge{l}}^{x} = L_{\edge{u}}^{x} = 0$. Likewise one finds $L_{\edge{l}}^{y} = L_{\edge{u}}^{y} = -L_{\edge{u}}^{y}$ (third column) and $L_{\edge{u}}^{y} = -L_{\edge{l}}^{y} = L_{\edge{l}}^{y}$ (fifth column), implying $L_{\edge{l}}^{y} = L_{\edge{u}}^{y} = 0$. For the $z$-component, however, we arrive at $L_{\edge{l}}^{z} = -L_{\edge{u}}^{z}$ (fourth and seventh column).

In summary, $L^{x}$ and $L^{y}$ vanish in the entire ribbon. In contrast,   $L^{z}$ is allowed nonzero but antisymmetric ($L_{\edge{l}}^{z} = -L_{\edge{u}}^{z}$) with respect to the central $x$-line of the nanoribbon; this means that the OAM in both regions cancel each other. These symmetries are fully confirmed in the simulations.

\section{IV. Snapshots of the dynamics}

\begin{figure}
    \centering
    \includegraphics[width=0.8\textwidth]{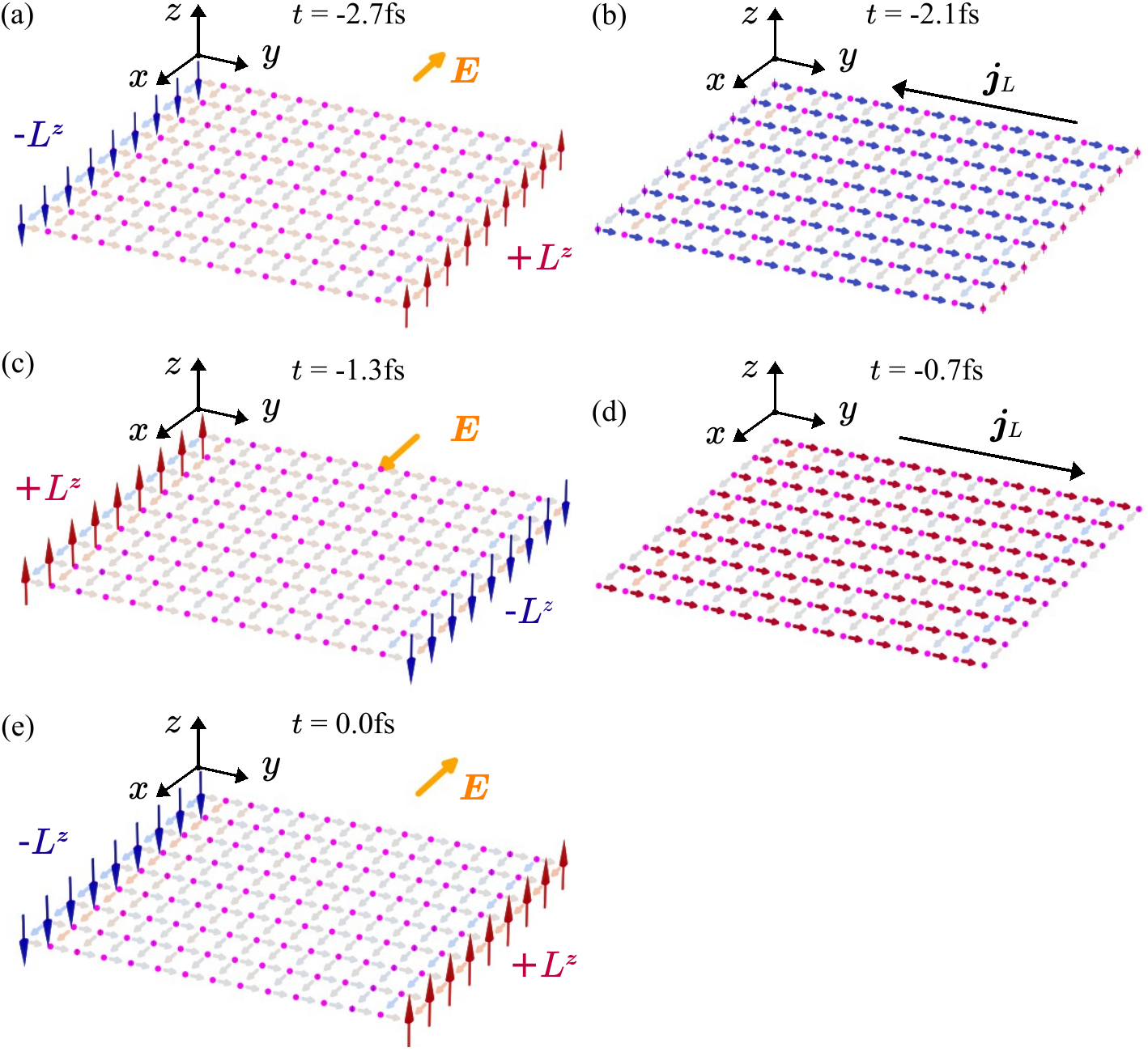} % for appendix in arXiv
    \caption{Snapshots of the orbital angular momentum (OAM) dynamics for a full period of the laser pulse. Small magenta spheres represent the Cu sites on a square lattice [fcc(001) monolayer]. Horizontally oriented small arrows along either $+x$- or $+y$-direction (connecting neighboring Cu atoms) represent the link direction and, thus, the direction of the OAM currents; their color encodes the magnitude and orientation of the $L^{z}$-polarized currents (orientation: red positive, blue negative, gray zero; magnitude: color saturation). Panels~a--e show situations at selected times $t$ (as indicated). Vertically oriented arrows in panels a, c and e display the induced OAM (red positive $L^{z}$, blue negative $L^{z}$.)}
    \label{fig:snapshots}
\end{figure}

In accordance with the periodic boundary conditions `along the ribbon', all quantities are $x$-translation invariant. Besides the probability current we find an $L^{z}$-current in $x$-direction, that is a longitudinal response to the electric field $E$. Focusing on transverse responses, the longitudinal OAM current is not discussed in this Paper.

The dynamics starting `before the laser pulse' and ending `after the pulse' may be studied by means of the animation included in the Supplemental Material. Here, we briefly discuss the dynamics within the laser's period before its maximum, that is from $t = -\unit[2.7]{fs}$ to $t = \unit[0]{fs}$. The data represented in Fig.~\ref{fig:snapshots} are those used in the main text.

\begin{itemize}
    \item Panel~a. At $t=\unit[-2.7]{fs}$, that is one period before the laser's amplitude maximum,  the linear polarized laser's electric field $\vec{E}$ is oriented along $-x$-direction (orange arrow) and OAM $\pm L_z$ (vertical red and blue arrows, i.e. in $\pm z$-direction) accumulates at the edges of the ribbon. OAM currents are small.

    \item Panel~b. A quarter period later ($t = \unit[-2.1]{fs}$, $E = 0$), transverse OAM currents $\vec{j}_L$ transport OAM across the ribbon in $-y$-direction (blue $L_z$-currents) and the OAM is strongly diminished at the edges.

    \item Panel~c. The situation half a period later ($t=\unit[-1.3]{fs}$) is  similar to that in panel~a,  but the reversed $E$ field (along $+x$) leads to opposite signs of the induced OAM\@.

    \item Panel~d. At $t = \unit[-0.7]{fs}$ the situation is  similar to that in (b), the direction of the OAM currents $\vec{j}_L$ is reversed. 

    \item Panel~e. One period later ($t=\unit[0]{fs}$) we find a situation very close to that in (a). The laser's amplitude is maximum.
\end{itemize}

\section{V. Classical analogon for the laser-driven dynamics}
Our simulations show that the orbital angular momentum accumulated at the sample's edges is reminiscent of an harmonic function of time~$t$. It is in phase with the electric field of the laser or has a phase shift of $\pi$, depending on the considered edge [Fig.~4(c) of the main text]. The transverse OAM currents are harmonic as well, but have their maximum a quarter of a period $T/4$ earlier [Fig.~3(d) of the main text]. The OAM currents are largest when the electric field vanishes. In addition, the transverse OAM currents exhibit the same phase relations for periodic boundary conditions in $y$-direction, where there are no edges and thus no accumulation, as is found in respective simulations (not shown here). 

These findings suggest that the OAM currents are primarily \emph{not} caused by equilibration of the accumulated OAM once the field is reduced or vanishes. Moreover, the time difference of $T/4$ cannot be understood within the scenario of a constant electric field, in which the transverse OAM current would be \emph{proportional to and in phase with} the electric field, as would follow from $j_{y}^{z} = \sigma_{yx}^{z} E_{x}$ ($\sigma_{yx}^{z}$ element of the OAM conductivity tensor).

In order to illuminate the above phenomena, we identify the accumulated OAM with the mechanical displacement $x(t)$ of a driven, damped harmonic oscillator and the OAM currents with its velocity $v(t) = \mathrm{d}x(t) / \mathrm{d}t$. In the respective Newtonian equation of motion
\begin{align}
    \frac{\mathrm{d}^2 x(t)}{\mathrm{d}t^2} + 2 \xi \omega_0 \frac{\mathrm{d}x(t)}{\mathrm{d}t} + \omega_0^2 x + A_0 \sin(\omega_{\mathrm{d}} t) & = 0,
\end{align}
$\xi$ and $\omega_0$ are the damping constant and the eigenfrequency of the undamped oscillator, respectively. The driving force with amplitude $A_{0}$ is considered harmonic, which allows to identify it with the electric field of the laser ($A_0 = e E_0 / m$). $\omega_{\mathrm{d}}$ thus mimics the laser's carrier frequency that is higher than the characteristic frequency of the electron's motion, which translates into $\omega_{\mathrm{d}} \gg \omega_0$. $\omega_0$ defines a characteristic timescale which is determined by the hopping parameters in the tight-binding Hamiltonian.

The complete solution is a superposition of a transient and a steady state. Due to damping, the transient becomes negligible at $t \to \infty$. The steady state 
\begin{align}
    x(t) & = x_0 \sin\left(\omega_{\mathrm{d}} t + \Delta\phi\right) \label{eq:steady_x}
\end{align}
is an harmonic oscillation with frequency $\omega_{\mathrm{d}}$ and amplitude
\begin{align}
    x_0 & = \frac{A_0}{\omega_{\mathrm{d}}\sqrt{(2\omega_0\xi)^2+\left(\frac{\omega_{\mathrm{d}}^2-\omega_0^2}{\omega_{\mathrm{d}}}\right)^2}}.
\end{align}
It is shifted in phase by
\begin{align}
    \Delta\phi & =\arctan\left(\frac{2\omega_{\mathrm{d}}\omega_0\xi}{\omega_{\mathrm{d}}^2-\omega_0^2}\right)
\end{align}
with respect to the driving force. In the limit $\omega_{\mathrm{d}} \gg \omega_0$ this becomes $\Delta\phi\rightarrow 0$. Likewise, the velocity
\begin{align}
    v(t) & = x_0 \omega_{\mathrm{d}} \sin\left(\omega_{\mathrm{d}} t + \Delta\varphi\right),\quad \Delta\varphi=\Delta\phi+\pi/2, \label{eq:steady_y}
\end{align}
is shifted in phase by $\Delta\varphi \rightarrow \pi/2$, which translates into a time shift by $T / 4$.

An exemplary solution with dimensionless parameters $\xi = 0.25$, $\omega_0 = 2 \pi$, $\omega_{\mathrm{d}} = 3\omega_0$, and $A_0 = 1$ is shown in Figs.~\ref{fig:harmonic_oscillator}a -- c. The transient becomes negligible at $t > 4$ (panels~b and c). Concerning the steady state, its velocity is maximum approximately when the driving field is zero (note that here $\omega_{\mathrm{d}}$ is only three times as large as $\omega_0)$. In the relevant regime $\omega_{\mathrm{d}} \gg \omega_0$, the phase shift becomes $\Delta \varphi \rightarrow \pi/2$ (panel~d) and the amplitude decreases (panel~e), as less and less energy can be `soaked up' by the oscillator.

\begin{figure*}
    \centering
    \includegraphics[width=0.8\textwidth]{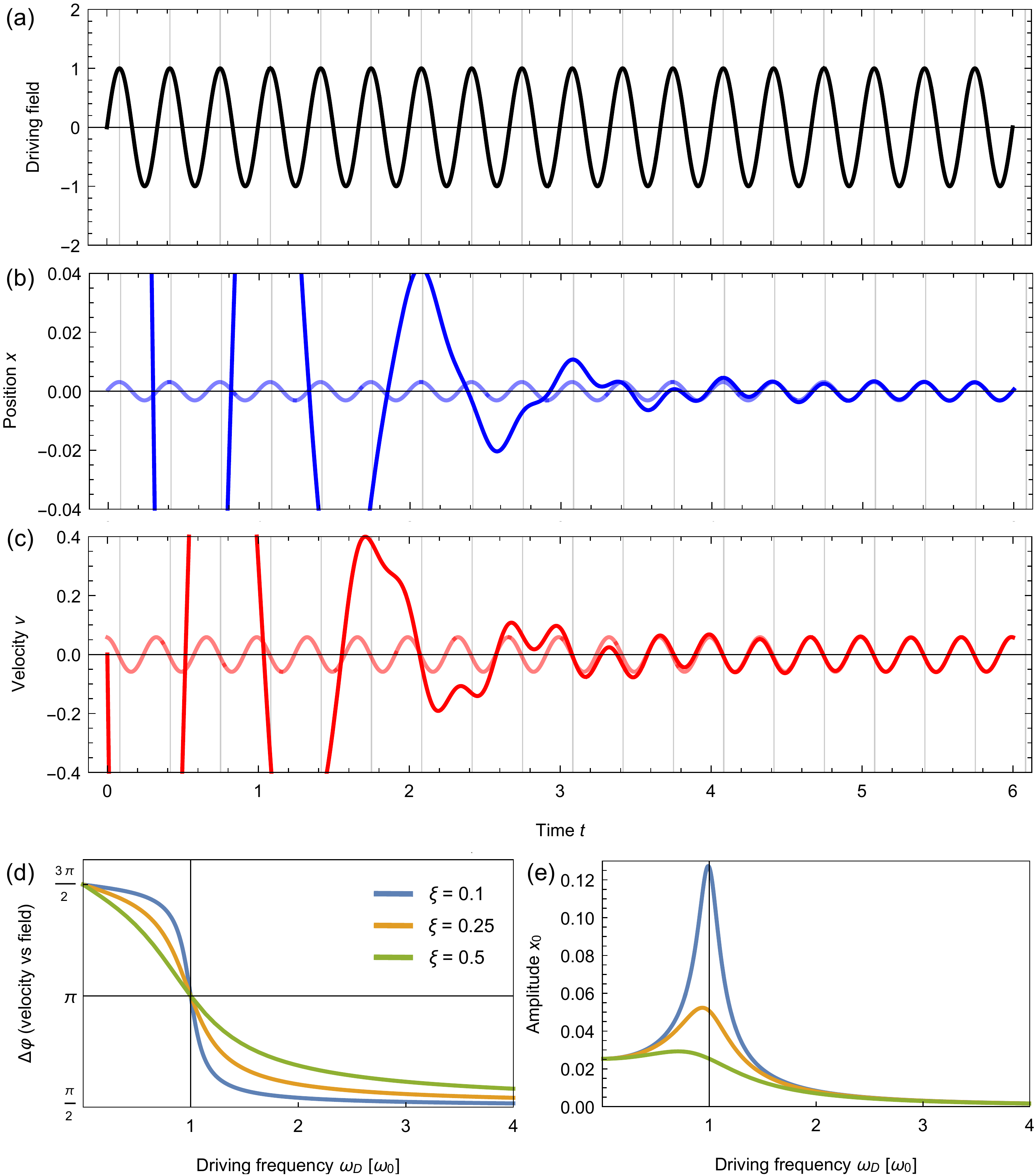}
    \caption{Driven harmonic oscillator. Panels (a) -- (c) depict the amplitude of the driving force, the position $x(t)$, and the velocity $v(t)$ versus time $t$, respectively. Pale blue and red curves show the steady-state solution according to equations~\eqref{eq:steady_x} and \eqref{eq:steady_y}. Panels (d) and (e) show the phase shift $\Delta \varphi$ of the velocity with respect to the driving force and the amplitude of the position $x_0$ of the steady-state solution for various damping strengths $\xi$ (as indicated in panel~d). The parameters are given in the text.}
    \label{fig:harmonic_oscillator}
\end{figure*}

We conclude that the analogon of a classical driven harmonic oscillator corroborates the ultrafast electron dynamics, since it exhibits the same phase relations as the OAM and OAM currents with respect to the electric field of the laser (confer Figs.~2a and 3d of the main text). Put another way, the ultrafast orbital Hall effect is actually laser-driven; it is not a time series of steady orbital Hall effects for varying electric field.
    
\end{widetext}

\end{document}